\documentclass[sigconf]{acmart}
\usepackage{graphicx} 
\usepackage{pifont}
\usepackage{xcolor}
\definecolor{BrickRed}{RGB}{203,65,84}     
\definecolor{ForestGreen}{RGB}{34,139,34}
\usepackage{booktabs}        
\usepackage{multirow} 
\usepackage{array}
\usepackage{enumitem}
\usepackage{tabularx}
\usepackage{tikz}

\usetikzlibrary{shapes.geometric}

\definecolor{QuestionColor}{HTML}{4F6D7A}    
\definecolor{SuggestionColor}{HTML}{7A9E7E} 
\definecolor{ModelColor}{HTML}{5DA9A8}      
\definecolor{VanillaColor}{HTML}{A398B3}    
\definecolor{ControlColor}{HTML}{D9CFC1}  
\definecolor{MIterColor}{HTML}{C9916B}   
\newcommand{\new}[1]{\textcolor{black}{#1}}

\newcommand{\questionbadge}{\tikz[baseline=-0.5ex]\node[draw=QuestionColor, fill=QuestionColor!20, rounded corners=2pt, inner sep=2pt, font=\tiny\bf] {Q};}
\newcommand{\suggestionbadge}{\tikz[baseline=-0.5ex]\node[draw=SuggestionColor, fill=SuggestionColor!20, rounded corners=2pt, inner sep=2pt, font=\tiny\bf] {S};}
\newcommand{\modelbadge}{\tikz[baseline=-0.5ex]\node[draw=ModelColor, fill=ModelColor!20, rounded corners=2pt, inner sep=2pt, font=\tiny\bf] {M};}
\newcommand{\vanillabadge}{\tikz[baseline=-0.5ex]\node[draw=VanillaColor, fill=VanillaColor!20, rounded corners=2pt, inner sep=2pt, font=\tiny\bf] {V};}
\newcommand{\controlbadge}{\tikz[baseline=-0.5ex]\node[draw=ControlColor, fill=ControlColor!20, rounded corners=2pt, inner sep=2pt, font=\tiny\bf] {C};}
\newcommand{\miterbadge}{\tikz[baseline=-0.5ex]\node[draw=MIterColor, fill=MIterColor!20, rounded corners=2pt, inner sep=2pt, font=\tiny\bf] {M-I};}

\AtBeginDocument{%
  }

\copyrightyear{2026}
\acmYear{2026}
\setcopyright{cc}
\setcctype{by}
\acmConference[CHI '26]{Proceedings of the 2026 CHI Conference on Human Factors in Computing Systems}{April 13--17, 2026}{Barcelona, Spain}
\acmBooktitle{Proceedings of the 2026 CHI Conference on Human Factors in Computing Systems (CHI '26), April 13--17, 2026, Barcelona, Spain}
\acmPrice{}
\acmDOI{10.1145/3772318.3791185}
\acmISBN{979-8-4007-2278-3/2026/04}

\begin{document}

\author{Sebastian Maier}
\email{maier.sebastian@campus.lmu.de}
\affiliation{
  \institution{LMU Munich \& Munich Center for Machine Learning (MCML)}
  \city{Munich}
  \country{Germany}
}

\email{}
\author{Manuel Schneider}
\affiliation{
  \institution{LMU Munich}
  \city{Munich}
  \country{Germany}
}

\author{Stefan Feuerriegel}
\affiliation{
  \institution{LMU Munich \& Munich Center for Machine Learning (MCML)}
  \city{Munich}
  \country{Germany}
}

\title{Partnering with Generative AI: Experimental Evaluation of Human-Led and Model-Led Interaction in Human-AI Co-Creation}

\begin{abstract}
Large language models (LLMs) show strong potential to support creative tasks, but the role of the interface design is poorly understood. In particular, the effect of different modes of collaboration between humans and LLMs on co-creation outcomes is unclear. To test this, we conducted a randomized controlled experiment ($N = 486$) comparing: (a) two variants of reflective, human-led modes in which the LLM elicits elaboration through suggestions or questions, against (b) a proactive, model-led mode in which the LLM independently rewrites ideas. By assessing the effects on idea quality, diversity, and perceived ownership, we found that the model-led mode substantially improved idea quality but reduced idea diversity and users’ perceived idea ownership. The reflective, human-led mode also improved idea quality, yet while preserving diversity and ownership. \new{We independently validated the findings in a different context ($N = 640$).} Our findings highlight the importance of designing interactions with generative AI systems as reflective thought partners that complement human strengths and augment creative processes.
\end{abstract}



\maketitle

\begin{figure*}[htbp]
    \Description[Study workflow across five experimental conditions]{Study workflow diagram with three labeled parts. Panel A shows five conditions: control (participant writes an initial idea only), vanilla chatbot (chat interface with a short exchange), model-led (participant provides an initial idea and the system rewrites it), suggestion-mode (system presents three suggestions), and question-mode (system asks reflective questions such as "What if" and "Explain"). An arrow leads to Panel B, where the participant submits a refined idea. Panel C shows evaluation of the refined ideas on three outcomes: idea diversity, idea quality, and perceived idea ownership.}

    \centering
    \includegraphics[width=0.9\textwidth]{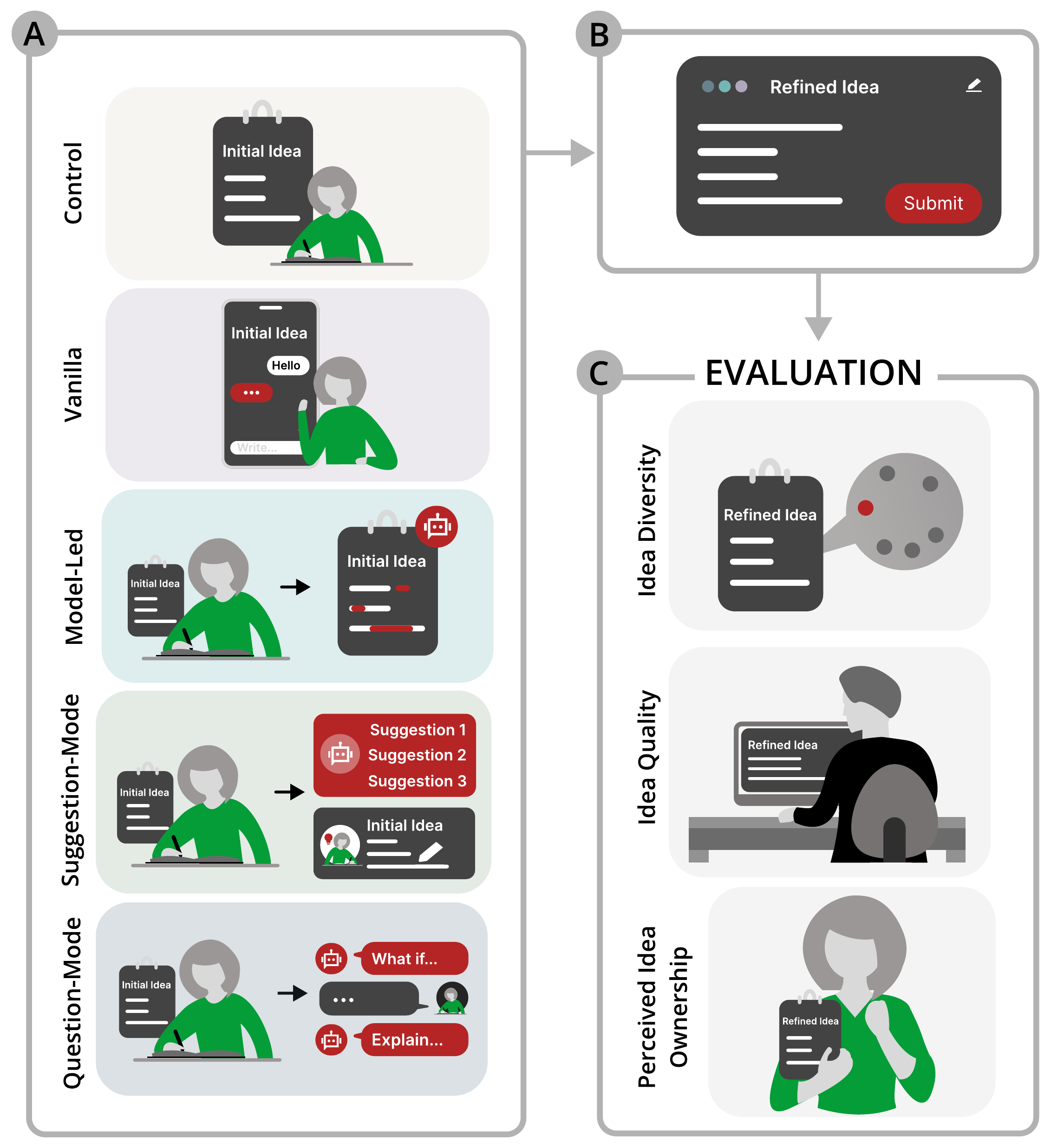}
    \Description[Study procedure with five conditions, refinement, and evaluation.]%
{The figure illustrates the study procedure. Panel A shows five experimental conditions in which participants first write down an initial idea. In the control condition, they receive no support. In the vanilla condition, the initial idea is directly entered into the system. In the model-led condition, the system requires additional input from the participant. In the suggestion-mode condition, the system provides multiple suggestions. In the question-mode condition, the system asks the user questions. Panel B shows that all participants can then refine their idea before submitting it. Panel C shows the evaluation stage, where ideas are assessed for diversity, quality, and perceived ownership.}

    \caption{\textbf{Overview of the study procedure of the main study.} \textbf{(A)} Participants ($N=486$) were randomly assigned to one of five conditions: four LLM variants (vanilla, model-led, suggestion, question) or a control group without support. \textbf{(B)} Before submitting their final ideas, all participants were instructed to refine their idea using the LLM. \textbf{(C)} Finally, idea quality was assessed by expert raters with domain expertise in the automotive industry; idea diversity was measured via text-embedding similarity of idea descriptions; and perceived ownership was self-reported by participants for the ideas they generated.
}
    \label{fig:studygraphics}
\end{figure*}

\section{Introduction}
\label{sec:Introduction}

Large language models (LLMs) have opened new opportunities for supporting creative work \cite{lee_empirical_2024, doshi_generative_2024, hubert_current_2024, feuerriegel_generative_2024}. The ability of LLMs to generate meaningful text enables writers to co-develop storylines, entrepreneurs to brainstorm novel business directions, and designers to expand early-stage ideas. Unlike earlier creativity support tools, LLMs allow for interaction in natural language, making them well-suited for iterative exploration and collaboration in creative processes.

Using LLMs as co-creation tools holds promise for boosting creative outcomes, yet the benefit of LLMs is contested due to three tensions. \emph{First}, there are ongoing discussions about whether LLMs are truly “creative” \cite{chakrabarty_art_2024}. Unlike humans, LLMs lack tacit knowledge and domain expertise, which are often essential for coming up with creative ideas of high \textbf{quality}, that is, ideas that are useful and contextually appropriate \cite{baer_importance_2015, lin_differential_2022}. Hence, co-creation systems are especially powerful when they complement, rather than replace, human capabilities via interactive collaboration \cite{rafner_creativity_2023}. \emph{Second}, if reliance on LLMs becomes widespread, the risk emerges that ideas will converge, reducing \textbf{diversity} across creative domains \cite{anderson_homogenization_2024, doshi_generative_2024, meincke_chatgpt_2025}. However, homogenization threatens one of the core characteristics of creativity---that is, novelty---and raises concerns about cultural and intellectual variety \cite{nakadai_ai_2023}. \emph{Third}, while LLMs might support the quality of creative output, their dominant role in shaping ideas may undermine users’ perceived \textbf{ownership} \cite{xu_what_2024, joshi_writing_2025, qin_timing_2025, draxler_ai_2024, li_value_2024, mirowski_co-writing_2023}. However, undermining users' perceived ownership is closely tied to lower user satisfaction and output valuation \cite{krauss_owning_2025}. Here, we argue that the three tensions can be addressed through careful interaction design by focusing on the way humans are involved in the process.

While existing studies on LLMs in creative contexts have examined the creativity abilities of LLMs (see Section~\ref{sec:Related_Work} and the overview paper in \cite{holzner_generative_2025}), the role of the interaction design is largely ignored. In particular, existing empirical studies focus on a simple interaction format (''one-shot``) \cite{holzner_generative_2025, hubert_current_2024, koivisto_best_2023}, where the input to the model is limited to a single prompt, so that iterative refinements are ignored. Yet, creativity theory stipulates that creativity is not a one-off act but an ongoing, cyclical process of generating, evaluating, and refining ideas \cite{wallas_art_1926, guilford_nature_1967, frich_mapping_2019}. Hence, it is natural that also LLMs for co-creation should be designed as iterative thought partners that scaffold human reflection rather than as autonomous creators \cite{rafner_creativity_2023}. Motivated by this, we stipulate that LLMs can be integrated in two main ways: (1) model-led, with the model taking initiative by rewriting or expanding ideas, or (2) human-led, with the model eliciting elaboration through questions or suggestions while keeping humans in control of the creative process. Yet, empirical evidence on how different human-led vs. model-led interaction designs and thus general HCI recommendations for such co-creativity systems are lacking. Hence, we explore how different designs may support perceived idea ownership, leverage idea quality, and preserve idea diversity in creative processes. 

\begin{quote}
\textbf{Research Question.}  
\emph{How does the interaction mode with LLMs (human-led vs.\ model-led) influence the quality, diversity, and ownership of ideas in co-creative tasks?}
\end{quote}

\noindent
To address this research question, we conducted a pre-registered randomized controlled experiment ($N = 486$). In particular, we compared five different conditions: (1) a \textbf{human-led question-mode}, where the LLM asked reflective questions that should trigger human thinking; (2) a \textbf{human-led suggestion-mode}, where the LLM provided tips to refine the idea;  (3) a \textbf{model-led mode} in which LLMs proactively refined participants’ ideas; (4) a \textbf{vanilla LLM}, with the default behavior of the LLM that was left unchanged; and (5) a \textbf{control group} without any LLM support. Participants were asked to generate creative ideas using an established task, namely, the Alternative Uses Test (AUT) \cite{guilford_nature_1967}, which involves repurposing existing objects (i.e., finding new applications for car sensors). Afterwards, ideas were evaluated along three dimensions: idea quality, idea diversity, and perceived idea ownership. \new{We independently validated our findings on LLM-enabled co-creation in the more naturalistic setting of product ideation (i.e., in a second, preregistered experiment with $N=640$ participants).}

Our research makes the following contributions:  
\begin{enumerate}
    \item Empirically, we provide the first large-scale controlled experimental evaluation (total $N=1,126$) directly comparing model-led and human-led interaction modes with LLMs in a complex creativity task, which shows that model-led idea refinement improves idea quality at the cost of idea diversity and perceived idea ownership. In contrast, actively involving participants with a question-driven interaction mode produces high-quality ideas while preserving ownership and even enhancing diversity.
    \item Theoretically, we identify active human involvement as a mitigation strategy for the quality--diversity trade-off of LLMs, thereby advancing HCI theory on effective human–AI co-creation. 
    \item Practically, our study suggests that designers of LLM co-creation systems should actively guide users in deciding when to delegate tasks to the AI and when to use the AI as scaffolding for their own thinking.
\end{enumerate}

\begin{table*}[htbp]
\centering
\caption{Research Gaps in Human-AI Co-creation Research.}
\label{tab:research-streams}
{\footnotesize
\begin{tabularx}{\textwidth}{p{0.16\textwidth}p{0.50\textwidth}X}
\toprule
\textbf{Research stream} & \textbf{What prior work shows} & \textbf{Open questions} \\
\midrule
\raggedright\textbf{Model capabilities for creativity} &
LLMs using one-shot prompts or default LLMs (like our vanilla bot) match or surpass average humans on established benchmarks for divergent thinking (e.g., Alternative Uses Test [AUT]) \cite{lee_empirical_2024, hubert_current_2024}. However, while LLM support boosts humans on average, the best humans still surpass LLMs in benchmark tests \cite{koivisto_best_2023}. &
How do LLMs perform in interactive co-creation processes that reflect real-world application scenarios? \\
\midrule
\raggedright\textbf{Designing systems for co-creation} &
Research on co-creation using AI can be loosely grouped into three main streams:
\begin{itemize}[leftmargin=6pt,itemsep=0pt,parsep=0pt,topsep=0pt]
\item \textbf{Artifacts/system contributions}: Tailored systems have been developed as creativity support tools, often using LLMs, such as for creative problem solving, collaborative songwriting \cite{kim_amuse_2025}, group ideation tools \cite{shaer_ai-augmented_2024}, or non-linear design processes \cite{zhou_understanding_2024}.
\item \textbf{Theoretical papers}: 
\new{This line of research outlines theoretical frameworks describing how creativity unfolds in co-creation with technology. It emphasizes that creativity unfolds in iterative, non-linear ways, and therefore advocates AI as a collaborative thought partner instead of a tool for simple command execution \cite{rafner_creativity_2023, zhou_understanding_2024, collins_building_2024}.} Further, being exposed to AI content during such creativity processes can lead to idea fixation \cite{wadinambiarachchi_effects_2024}.
\item \textbf{Experimental evaluations}: Some design choices are studied experimentally, such as, the timing (who proposes the idea first, i.e., human or AI) \cite{qin_timing_2025}.
\end{itemize} &
What are effective interaction modes, where creativity emerges in the collaboration with the LLM itself? \\
\midrule
\raggedright\textbf{Tensions in human-AI co-creation} &
Prior work shows several tensions that arise in the co-creation process: 
\begin{itemize}[leftmargin=6pt,itemsep=0pt,parsep=0pt,topsep=0pt]
\item \textbf{Decrease in idea quality}: People perform worse unaided after interacting with the AI than before \cite{kumar_human_2025}.
\item \textbf{Homogenization effect}: Interaction with LLMs (and LLMs alone) reduces the diversity of the idea pool \cite{doshi_generative_2024, anderson_homogenization_2024, kadoma_role_2024}.
\item \textbf{Psychological side-effects}: Co-creating with AI reduces psychological ownership \cite{joshi_writing_2025, draxler_ai_2024}.
\end{itemize} &
How can we address these tensions? \\
\bottomrule
\end{tabularx}
}
\end{table*}

\section{Related work}
\label{sec:Related_Work}

\subsection{Conceptualization of Creativity }
\label{sec:ConceptCreativity}

Creativity is a multidimensional phenomenon, which is commonly understood as the production of ideas or solutions that are both original and appropriate to a given task \cite{kaufman_cambridge_2019}. Originality reflects how far an idea diverges from conventional thinking. It often emerges when prior knowledge is connected in novel ways (for example, through analogical reasoning, which leads to outputs that feel new and unexpected \cite{plucker_psychometric_1999}. Appropriateness refers to an idea’s feasibility and usefulness, that is, whether it can realistically address a problem and improve the given context \cite{sternberg_concept_1999, runco_creativity_2005}. Originality and appropriateness are typically assessed at the individual level. 

In addition, at the collective level, the diversity of the idea pool can also be relevant in practice \cite{anderson_homogenization_2024}, which reflects how distinct the ideas are from one another within a group or condition. Importantly, greater diversity is relevant in practice because it expands the solution space and increases the likelihood of identifying novel and valuable directions. Beyond the above outcome-oriented perspectives, research has also emphasized the subjective experience of the creative process. For example, the extent to which individuals experience perceived ownership over an idea could shape how they are committed to it \cite{gray_emergence_2020}.  In sum, the creative process is commonly evaluated along several dimensions, namely: (i) idea quality (originality and appropriateness), (ii) the diversity of generated ideas, and (iii) the subjective experience of the process, such as perceived ownership.

\subsection{Augmenting Human Creativity}
\label{sec:Creativity}

In HCI, scholars have long been interested in supporting human creativity through technology, often subsumed under the umbrella of creativity support tools (CSTs) \cite{shneiderman_creativity_2002}. CSTs have been applied across various domains, including visual design, storytelling, and ideation platforms \cite{frich_mapping_2019}.  Previous creativity research in HCI has evolved in waves with two peaks in the 1990s (individual tools) and 2000s (collaborative platforms) \cite{frich_mapping_2019}, while a third wave is seen in response to the advances of large language models, where technology might have reached a stage where it can, to some extent, exhibit creativity on its own. Here, creativity is typically assessed through standardized tests such as the Alternative Uses Test (AUT), which asks to generate as many novel and diverse uses as possible for a common object, such as a brick \cite{guilford_nature_1967}. In these benchmark tests, LLMs are very powerful and can even surpass average human performance \cite{hubert_current_2024}. \new{However, in more complex creative tasks, such as creating a new innovative product for a start-up or writing novels, LLMs remain less capable on their own \cite{holzner_generative_2025}. \citet{kumar_human_2025} empirically show that exposure to LLM-generated ideas and strategies can increase performance during creativity benchmarks but can hinder later unaided generation. However, the assistance in their study was static and thus without interactive elements: participants viewed pre-generated lists of ideas or cues rather than engaging in a dynamic dialogue with an LLM, leaving open how interactive co-creation might shape creative performance. HCI research and theory, such as the Co-Creative Framework for Interaction (COFI) \cite{rezwana_designing_2023}, emphasize the relevance of the specific interaction design to create effective co-creation. Still, there is no experimental evaluation of whether and how interactive collaboration between humans and LLMs can augment creativity in such settings.}

From research in social psychology, we know that creativity emerges from social interactions in the moment, where people listen, respond, and build on one another \cite{sawyer_individual_2019}. Like in jazz, one instrument starts to play and others improvise on the spot, creating a completely new tune on the spot. Similarly, in the innovation context, new products might emerge in an exchange of people with different experiences and backgrounds, like designers, computer scientists, and economists. \new{Creativity support design principles emphasize the relevance of exploratory, collaborative, and iterative processes in which ideas are continuously expanded, adapted, and refined \cite{resnick_design_2005}.} However, this iterative and collaborative process remains underexplored in current HCI research on LLMs and creativity. 

\subsection{Interaction Modes: Model-Led vs. Human-Led}

Designing human-AI interaction modes in a way that supports human capabilities is highly challenging \cite{vaccaro_when_2024, senoner_explainable_2024}. For example, in the context of decision making, AI often constrains humans as they do not know when to trust the model. For creativity, LLMs on their own are often as effective as when paired with humans \cite {lee_impact_2025}. However, this is associated with negative effects, such as reduced idea diversity and psychological ownership \cite{doshi_generative_2024, anderson_homogenization_2024}. Thus, effective creative collaboration between humans and LLMs remains challenging.

Psychological theory offers ways to formalize the challenges in human collaborative settings: \emph{social loafing} refers to the tendency of individuals in a group to contribute less effort compared to when they work alone \cite{latane_many_1979, karau_social_1993}. For example, in a brainstorming session, individuals may put less effort into generating novel ideas when working in a large group compared to working alone \cite{diehl_productivity_1987, ez-zaouia_group_2024}. \new{In the context of suggestions for writing, individuals often prefer the low-effort support of selecting between suggestions compared to actively prompting the suggestions \cite{dang_choice_2023}}. Psychological theory suggests that when individuals lack control over an outcome, they may become demotivated and allocate fewer cognitive resources \cite{latane_many_1979, maier_learned_1976}. Translated into human–AI collaboration, this tendency can manifest as \emph{algorithmic loafing}: rather than engaging fully, individuals reduce their own effort and adopt the system’s output with little reflection \cite{inuwa-dutse_algorithmic_2023}. Applied to our creativity context, this means that, instead of actively striving for the best solution, people may disengage and settle for whatever the AI produces.

As a consequence, in the context of co-creating with LLMs, it might be beneficial to keep humans engaged and in charge of the creative process. Co-creation with LLMs often reduces human effort and makes the task less demanding \cite{mei_if_2025}. However, research shows that the mode of collaboration affects creative performance. When people co-create with a model rather than merely edit its output, they report higher self-efficacy, which, conversely, leads to higher creative quality \cite{mcguire_establishing_2024}. \new{This aligns with findings that co-creative AI systems foster higher engagement when users perceive them as collaborative partners that inspire new directions rather than as tools that merely execute commands \cite{lawton_when_2023}}.  Consequently, effective co-creation keeps humans involved in the creative process and triggers their creativity rather than allowing mental offloading. 

Extending this perspective, creativity research provides support on how to foster creative processes. \new{For example, guidance that prompts designers to actively reflect on their decisions can reduce over-reliance on AI-generated outputs \cite{gmeiner_exploring_2025}}. Additionally, individuals produce more novel ideas when they are provided analogies from unrelated domains that are still structurally relevant \cite{dahl_influence_2002}. These trigger cognitive processes and allow for producing novel content. CSTs might support here, for example, by providing analogies that are distinct but connected \cite{gilon_analogy_2018}. Building on this insight, we argue that human–LLM interaction modes are more effective when they stimulate the human creative process rather than letting the model take the creative lead. For example, when exploring new product ideas, individuals may either become constrained by the model's output or be inspired by novel analogies they might not have considered otherwise. Here, we distinguish between (1) \emph{model-led collaboration}, where the LLM generates the core ideas and the human mainly edits or reviews, and (2) \emph{human-led} collaboration, where the human drives ideation and the LLM supports through providing analogies, useful information, and supporting in elaboration. 

\new{Later, we adopt two different human-led interaction modes commonly used in HCI and creativity research: (i)~a \emph{suggestion mode}, where the AI offers alternative ideas, reflects a well-established pattern in creativity support systems that use suggestions to inspire user thinking \cite{dang_choice_2023, kumar_human_2025, doshi_generative_2024}, and (ii)~a \emph{question mode}, where the AI asks reflective questions, draws on work in HCI that employs socratic or reflection-based prompts to encourage users to articulate and extend their ideas \cite{winkler_sara_2020, gmeiner_exploring_2025, bastani_generative_2025, stenning_socratic_2016, al-hossami_socratic_2023}. Both interaction modes were designed to stimulate the participants' thinking and spark their creativity.}


\subsection{Research Gap}
In summary, prior research on LLMs lets us hypothesize that the interaction mode between humans and LLMs is relevant for the creativity outcomes; however, current studies do not test this empirically. As a result, it is unclear which interaction modes---in particular, whether the human or the model leads the creative process---are effective in promoting the idea quality, diversity of the ideas, and perceived ownership of ideas.

\section{Hypotheses}

\subsection{Individual Level: Idea Quality}

Co-creation with LLMs can help find ideas that are more original and appropriate. There is meta-analytical evidence that human-LLM co-creation leads to consistent positive effects \cite{holzner_generative_2025}. For example, LLMs support individuals in more creative and better-written short stories \cite{doshi_generative_2024}, creating ideas for innovation \cite{mei_if_2025}, or enhancing group brainwriting \cite{shaer_ai-augmented_2024}. We therefore hypothesize that LLM support, regardless of the interaction format, increases idea quality:

\begin{description}
    \item [H1:] \emph{Idea quality is higher when participants interact with (a) a \textbf{human-led question-mode} LLM, (b) a \textbf{human-led suggestion-mode} LLM, (c) a \textbf{model-led} LLM, or (d) a \textbf{vanilla} LLM compared to a control condition without LLM support.}
\end{description}

\subsection{Group Level: Idea Diversity}

While co-creation with LLMs might augment creativity on an individual level, this may still reduce diversity at the group level. Ideas generated by LLMs are more similar to each other compared to ideas generated by a human crowd, implying a possible homogenization effect \cite{meincke_chatgpt_2025}. Also, when humans collaborate with LLMs, the homogenization effect applies, so that ideas (or other text such as stories) may become more similar to each other and thus less diverse \cite{doshi_generative_2024,meincke_using_2024}. A reason for this effect can be human bias, such as fixation effects \cite{kohn_collaborative_2011, sio_fixation_2015, crilly_creativity_2019}, which also emerge in human-to-human collaborative settings. In contrast to this, we suggest that this effect could be mitigated by interaction modes, where the humans do the main creative work and the model rather acts as a source of inspiration compared to model-led conditions:

\begin{description}
    \item [H2:] \emph{The diversity of the idea pool is higher when participants interact with a \textbf{human-led question-mode} LLM compared to (a) a \textbf{model-led} LLM and (b) a \textbf{vanilla} LLM.}
    \item [H3:] \emph{The diversity of the idea pool is higher when participants interact with a \textbf{human-led suggestion-mode} LLM compared to (a) a \textbf{model-led} LLM and (b) a \textbf{vanilla} LLM.}
\end{description}

\subsection{Ownership of Outputs When Interacting With LLMs}

Collaboration with LLMs often reduces perceived ownership of ideas \cite{draxler_ai_2024, qin_timing_2025, joshi_writing_2025}. For example, when humans only provide a short prompt to LLMs to generate an idea for a competition, they might not perceive the idea as their own. However, in the context of innovation, this could have detrimental effects, because people need proper motivation and the will to follow along to bring ideas into practice \cite{berg_getting_2021, van_dyne_psychological_2004}. This effect can potentially be diminished by increasing human contribution \cite{joshi_writing_2025}. We therefore hypothesize:

\begin{description}
  \item[H4:] \emph{Perceived idea ownership is higher when participants interact with a \textbf{human-led question‑mode} LLM than with (a) a \textbf{model‑led} LLM and (b) a \textbf{vanilla} LLM.}
  \item[H5:] \emph{Perceived idea ownership is higher when participants interact with a \textbf{human-led suggestion‑mode} LLM than with (a) a \textbf{model‑led} LLM and (b) a \textbf{vanilla} LLM.}
\end{description}

\section{Methods}
\label{sec:Methods}
To empirically assess our hypotheses, we performed a preregistered (\url{https://aspredicted.org/nf5r-cmrt.pdf}) randomized controlled experiment with $N = 502$ participants. Codes and anonymized data are available via our GitHub (\url{https://github.com/SM2982/replication-creativity-CHI2026}). Our study compared human-led idea generation (question-mode and suggestion-mode) and model-led idea generation, along with a vanilla LLM condition and a no-LLM control condition. Our study received IRB approval from Ethics Committee at LMU Munich School of Management (ETH-SOM-004, approved March 12, 2025). All participants provided their informed consent and could withdraw at any time. Data were collected and stored in accordance with local data privacy laws. Compensation was done via Prolific, meeting the fair‑wage policy of the platform.

\begin{figure*}[htbp]
    \Description[Interfaces for question-mode and suggestion-mode]{Two interface screenshots with four callouts. Left side shows question-mode as a chat conversation. Callout A highlights a system message that adds domain knowledge about a vehicle feature and asks a reflective question. Callout B highlights a second system message using an analogy (for example, smart home systems) to prompt an alternative idea, followed by the participant response. Right side shows suggestion-mode. Callout C highlights a panel listing three written suggestions for improving novelty and usefulness. Callout D highlights a text box where the participant rewrites their idea based on the suggestions and then finishes.}

    \centering
    \includegraphics[width=1.0\textwidth]{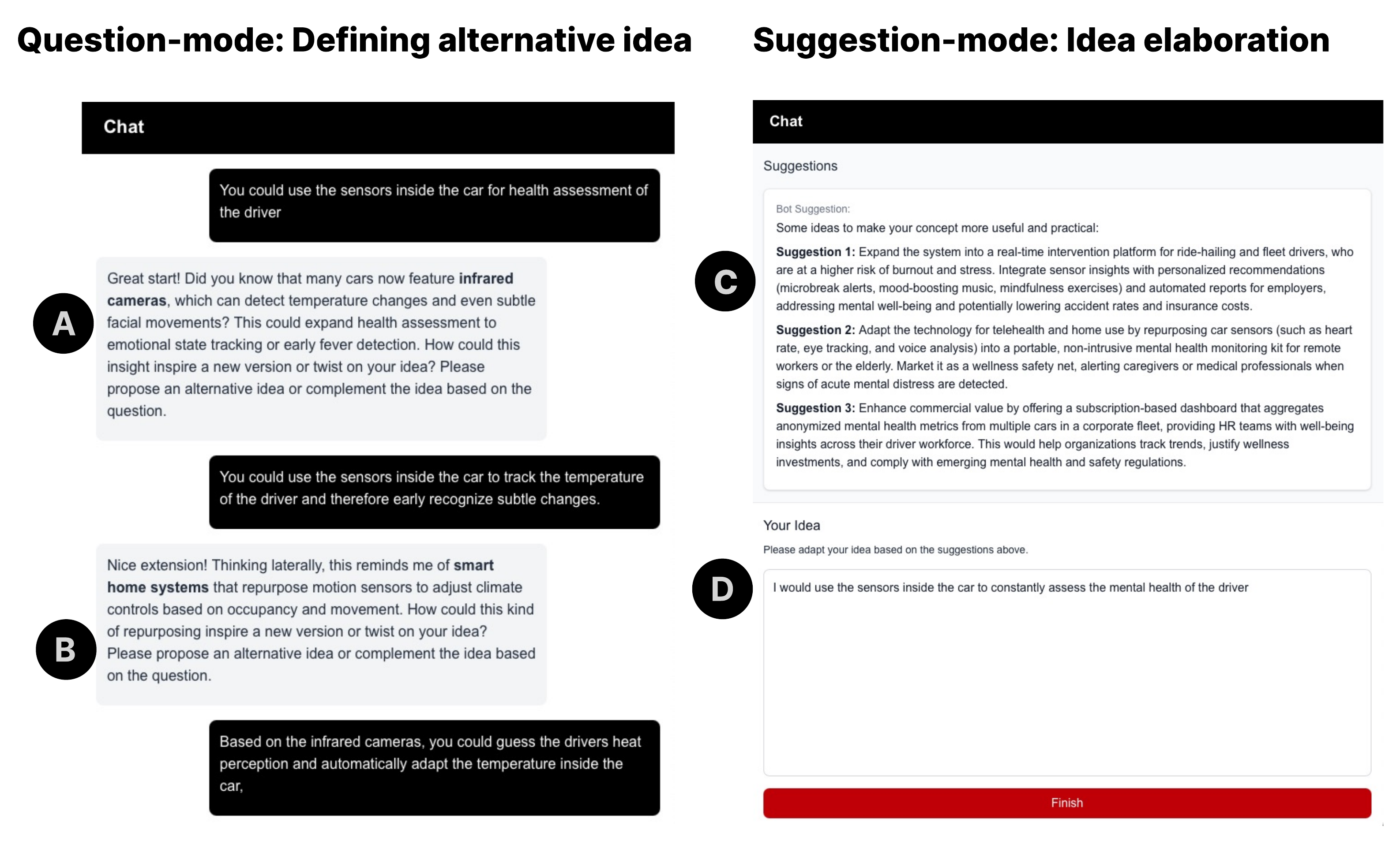}
    \Description[Interfaces of question-mode and suggestion-mode with example dialogues.]%
    {The figure shows screenshots of the two human-led conditions. On the left, the question-mode presents a chat-style interface. Panel A shows the system adding knowledge about infrared cameras to inspire alternative ideas. Panel B shows the system providing an analogy to smart home systems to stimulate further creativity. On the right, the suggestion-mode displays three written suggestions that propose ways to expand the initial idea, including intervention, telehealth use, and commercial applications. The text window beneath allows the participant to rewrite and refine the idea based on the suggestions.}
    \caption{\textbf{Screenshots of the interface of the two human-led conditions.} In the \textbf{question-mode}, the LLM provided \textbf{(A)} further knowledge and \textbf{(B)} an analogy to inspire human creativity. After defining three alternative ideas, the user chooses one to elaborate further. In the \textbf{suggestion-mode}, the participants were provided three suggestions on how to make the initial idea more novel, and on how to further elaborate \textbf{(C)}. Below the suggestion, a text box is shown where the participants can refine their initial idea based on the LLM input \textbf{(D)}.
}
    \label{fig:Screenshots}
\end{figure*}

\subsection{Procedure}
\label{sec:Procedure}
We recruited participants through Prolific and redirected them to a custom-built study website accessible on both computers and mobile devices (see Figure~\ref{fig:Screenshots}). Our study website combined a Next.js/React frontend (TypeScript, Tailwind CSS) with a FastAPI backend in Python. This architecture ensured experimental control, a highly interactive interface, and short response latencies to support a seamless participant experience. 

After providing informed consent, participants were randomly assigned to one of the five experimental conditions. In each condition, the participants were provided with an adapted version of the AUT task \cite{guilford_nature_1967}. That is, participants had to come up with a creative way to repurpose existing features of cars, such as steering wheels, brakes, cameras, sensors, screens, lights, or speakers---for entirely new uses beyond the original purpose in vehicles. We chose this task to avoid the risk of public benchmarks of training data contamination, which can reflect model performance via memorization rather than generalization. Furthermore, it is an ecologically valid and complex task that is used exactly like this in companies. We therefore focused our task on combining novelty with usefulness in an innovation context. 

An overview of the study procedure is shown in Figure~\ref{fig:studygraphics}. After participants were provided the task, participants either received support in an LLM condition or formulated the idea on their own. Before submitting the final idea, participants in all conditions could refine the final idea themselves. Afterwards, participants filled out questionnaires with different measurements (see details below). 

\subsection{Intervention Groups}
\label{sec:Intervention}
Participants in both the model-led and human-led conditions first generated an initial idea without any LLM assistance. They then engaged with the chatbot according to their assigned condition. To ensure comparability, all LLM-assisted conditions used GPT-4.1 via the OpenAI API (snapshot gpt-4.1-2025-04-14). We followed best practices for the prompt design in the model-led and human-led condition \cite{openai_gpt-41_2024}. The full list of prompts is provided in Appendix~~\ref{sec:prompts}. In total, we implemented five conditions. \new{In all model-led and human-led conditions, the intervention aimed to (i) enhance the idea's originality (exploration) and (ii) enhance its development (elaboration).} 

\begin{itemize}
    \item \textbf{Model-led LLM (\modelbadge)}: Participants provided an initial idea (max.\ 10 words). \new{By imposing a word limit, we ensured that the LLM-model performed the primary creative elaboration rather than merely refining an already developed idea}.\footnote{\new{In the validation study, we implemented a different word limit (max. 15 words) that is the same for all conditions, including both human-led and model-led conditions. Still, we arrived at the same conclusions (see Section~\ref{sec:Validation}).}} \new{A one-shot prompt was used to enhance originality and usefulness, after which participants could refine the model's output before submitting.}
    
    \item \textbf{Human-led LLM}: The model facilitated iterative elaboration under a reflective paradigm in two phases: First, in the novelty phase, participants generated alternative ideas to increase novelty, while, afterwards, in an elaboration phase, participants further developed the selected idea to increase appropriateness. Here, we tested two variants of the human-led condition:
    \begin{itemize}
        \item \textbf{Suggestion-mode} (\suggestionbadge): The LLM generated three alternative ideas in the first phase and three elaboration suggestions in the second phase. Participants refined the idea independently.
        \item \textbf{Question-mode} (\questionbadge): The LLM generated three alternative ideas in the first phase using cross-domain analogical reasoning \cite{ding_fluid_2023, dahl_influence_2002} and knowledge of car sensor functions. After participants selected one idea, the model asked targeted questions in the second phase to develop the idea further.
    \end{itemize}
 
    \item \textbf{Vanilla LLM} (\vanillabadge): Same chat interface as in the question-mode, but without additional system prompts or intervention logic. Participants could freely chat with GPT 4.1 and formulate or copy and paste the idea they generated. 
    
    \item \textbf{Control} (\controlbadge): Participants generated and refined their idea without any LLM support.
\end{itemize}

\noindent
In the question-mode, we implemented \emph{function calling}---a structured API mechanism enabling the experimental platform to trigger specific model actions (e.g., generating analogical alternatives, asking elaboration questions, switching phases). This ensured that interactions systematically followed the intended protocol and captured the idea in the process properly, which would not be possible with a single prompt. 

\new{We validated the prompt manipulation through iterative testing. First, we conducted multiple short sessions with external testers to verify that each prompt condition consistently elicited the intended interaction patterns and that instructions were clearly understood. We iteratively refined prompts based on observed deviations from intended behavior. Second, we conducted a pilot study with 50 participants to validate the complete technical system, confirm that function calls were executed as designed, and ensure participants understood the different interaction modes. We analyzed all dialogues and adapted the prompts and code to address minor issues.}

\subsection{Sample}
\label{sec:sample}

\new{We approximated our required sample size using G*Power for a one-tailed two-sample $t$-test. To detect at least a medium effect size ($d$ = 0.5) with 80\% power and $\alpha = 0.0125$ (Bonferroni correction as a conservative proxy for Holm), we needed approximately 78 participants per group. To account for potential exclusions such as due to attention checks, we aimed for 100 participants per group as preregistered.} Overall, \emph{N} = 502 participants from the U.K. were recruited via the platform Prolific. Inclusion criteria for our experiment were a minimum age of 18, English as a first language, an approval rate of at least 98\%, and at least 50 previous submissions. The final sample consisted of participants with an age range of 18 to 78 years (\emph{M} = 41.34, \emph{SD} = 12.99); 49.21\% identified as female, 50.61\% as male, and one participant preferred not to disclose their gender. Following our preregistered exclusion criteria, we removed participants who failed at least one attention check (10 participants), acknowledged unauthorized AI use during the study (2 participants), provided nonsense answers (2 participants), participated more than once (1 participant), or did not finish the questionnaire (1 participant). This resulted in a final sample of \emph{N} = 486 participants.

\subsection{Measures}
\label{sec:measures}

We focus on the following three measures:
\begin{enumerate}
    \item \textbf{Idea quality:}
To assess idea quality, we conducted the Consensual Assessment Technique (CAT), which is considered the gold standard to assess creativity \cite{amabile_social_1982}. As recommended, four domain experts---each with at least five months of experience in the innovation department of a major automotive manufacturer---rated all ideas independently on an originality and usefulness scale (the scale was adapted from \citet{lee_empirical_2024}). For each idea, we averaged the originality and usefulness ratings across all raters to obtain a single idea quality score. The consistency between the raters reached an ICC (3,k) = .60 and is thus considered ``good'' according to \citet{cicchetti_guidelines_1994}.

    \item \textbf{Diversity of the idea pool:} We generated embeddings with OpenAI’s \texttt{text-embedding-3-large} model and, in line with \citet{feuerriegel_using_2025}, measured per-idea diversity as the cosine distance to the leave-one-out centroid (mean of all other embeddings) within the same experimental condition. \new{This method has been found to align well with human judgments of conceptual diversity in previous studies \cite{anderson_evaluating_2024, beaty_automating_2021, dumas_measuring_2021}.}
    \item \textbf{Perceived idea ownership:} We measured perceived idea ownership with five items adapted from \citet{qin_timing_2025}. 
Participants indicate their ratings on a 7-point Likert scale (i.e., ``I have made substantial contributions to the content of the resulting idea''). The items showed an internal consistency of $\alpha $ = .87.
\end{enumerate}
Beyond the primary outcomes, we examined additional variables as part of an exploratory analysis. To measure \textbf{perceived cognitive workload}, we adapted two items from the NASA-TLX \cite{hart_nasa-task_2006}, namely, perceived mental demand and effort. Participants provided their ratings on a 0--100 slider scale (e.g., ``How mentally demanding was the task?''). The two items were strongly correlated ($r$ = .77). We further adopted three \textbf{user experience}-related scales, namely, performance expectancy, effort expectancy, and hedonic motivation (i.e., ``I find the chatbot useful for creative tasks.''). These scales were adopted from the UTAUT questionnaire \cite{venkatesh_consumer_2012} and internal consistency for the scales ranged from $\alpha $ = .92 to .97. We also assessed creative \textbf{self-efficacy} via three items adopted from \citet{tierney_creative_2002} (i.e., ``I was good at coming up with novel ways to repurpose vehicle features''; $\alpha $ = .89). Additionally, we performed an inductive qualitative content analysis \cite{hsieh_three_2005} to assess the participants' elaboration depth in their interactions. Drawing on Guilford’s construct of elaboration \cite{guilford_nature_1967}, as operationalized in the Torrance Tests of Creativity \cite{torrance_torrance_1966}, we treated elaboration as a sensitizing concept and transferred it to our interaction context. The first author open-coded the idea-refinement turns and, upon discussion with the other authors, iteratively organized incidents into seven characteristics for idea elaboration (CIEs) (i.e., adding a guiding principle, mechanism, feature, constraint, audience/context, integration, or reframing). Using these CIEs as decision rules, we derived a five-level ordinal scale of \textbf{elaboration depth}, ranging from Level 1 (AI-led echo, no new input) to Level 5 (human-led reframing, a radical pivot of the idea’s purpose or audience). This scheme allowed us to qualitatively assess the depth and originality of user contributions beyond simple acceptance of LLM output. \new{The detailed coding scheme can be found in the Appendix ~\ref{sec:elaboration-coding}}. The \textbf{demographic measures} were provided by the {Prolific} platform. For the context of our study, age and gender were included as demographic context variables in our analysis. Full measurement details, including all exploratory instruments and item wordings, are provided in the public GitHub repository.

\subsection{Analysis}
We analyzed each dependent variable using linear models with the condition as a predictor. Our preregistered one-sided hypotheses were tested via contrasts of estimated marginal means. To account for multiple comparisons across conditions, we adjusted $p$-values using the Holm method \cite{holm_simple_1979}, thereby controlling the family-wise error rate. Statistical significance for all hypothesis tests was set at $\alpha = .05$.  Effect sizes are reported as Cohen’s $d$ \cite{cohen_statistical_2013}. In addition, each estimate is reported alongside its $95\%$ confidence interval~(CI). All analyses were conducted in \emph{R} 4.4.1, using the \texttt{emmeans} package \cite{lenth_emmeans_2025} for calculating the contrasts. \new{As dependent variables included subjective ratings and bounded measures (cosine similarity) we conducted non-parametric robustness checks for all dependent variables (see Appendix~\ref{sec:RobustnessChecks}).}

\begin{table*}[p]
\centering
\caption{Descriptive Statistics for Key Measures Across Experimental Conditions}
\label{tab:condition_measures}
\footnotesize
\begin{tabular}{lc *{10}{c}}
\toprule
& & \multicolumn{2}{c}{Idea quality}
& \multicolumn{2}{c}{Idea diversity}
& \multicolumn{2}{c}{Perceived} 
& \multicolumn{2}{c}{Perceived}
& \multicolumn{2}{c}{Creative self-efficacy} \\
& & & & &
& \multicolumn{2}{c}{Ownership}
& \multicolumn{2}{c}{Cognitive load}
& \multicolumn{2}{c}{} \\
\cmidrule(lr){3-4} \cmidrule(lr){5-6}
\cmidrule(lr){7-8} \cmidrule(lr){9-10} \cmidrule(lr){11-12}
Condition & \textit{N} & Mean & SD & Mean & SD
& Mean & SD & Mean & SD & Mean & SD \\
\midrule
\questionbadge{} Question Mode  & 96  & 2.96 & 0.78 & 0.379 & 0.079 & 5.71 & 1.08 & 60.00 & 22.02 & 5.09 & 1.40 \\
\suggestionbadge{} Suggestion Mode& 95 & 2.78 & 0.63 & 0.333 & 0.104 & 5.24 & 1.14 & 58.26 & 22.18 & 4.70 & 1.29 \\
\modelbadge{} Model-Led      & 101 & 3.11 & 0.74 & 0.309 & 0.059 & 5.03 & 1.24 & 40.94 & 21.89 & 4.63 & 1.40 \\
\vanillabadge{} Vanilla        & 95  & 2.70 & 0.87 & 0.378 & 0.103 & 4.99 & 1.48 & 49.97 & 22.88 & 4.68 & 1.44 \\
\controlbadge{} Control        & 99  & 2.68 & 0.85 & 0.343 & 0.108 & 6.11 & 1.03 & 57.37 & 21.81 & 4.79 & 1.49 \\
\bottomrule
\end{tabular}

\bigskip
\raggedright\footnotesize
\emph{Note:} All measures are reported as means and standard deviations.
Idea diversity refers to within-condition similarity measures.

\end{table*}

\section{Results}
\label{sec:results}

\subsection{Hypothesis Testing}

\subsubsection{Idea Quality (H1): Significantly Higher for Question-Mode and Model-Led Conditions}

Hypothesis H1 predicted higher idea quality when participants interacted with (a) a question‐mode LLM, (b) a suggestion‐mode LLM, (c) a model‐led LLM, or (d) a vanilla LLM, compared to a control condition without LLM support. Distribution of the different conditions regarding idea quality is shown on the left-hand side of Figure~\ref{fig:violin-main}. Participants interacting with the question-mode (H1a: $t(481) = 2.49$, $p = .020$, $d = 0.36$) and the model-led mode (H1c: $t(481) = 3.86$, $p < .001$, $d = 0.55$) produced significantly higher quality ideas compared to the control condition without support. In contrast, participants interacting with the suggestion-mode LLM (H1b: $t(481) = 0.91$, $p = .366$, $d = 0.13$) and the vanilla LLM (H1d: $t(481) = 0.13$, $p = .449$, $d = 0.02$) did not produce significantly higher quality ideas compared to the control condition without support. Therefore, Hypothesis H1 is \textbf{partially supported}: the question-mode and model-led modes demonstrated significantly higher idea quality over the control condition, while the suggestion and vanilla models showed no significant improvement over baseline.

\subsubsection{Idea Diversity (H2\&H3): Significantly Higher for Question-Mode Condition}
Hypotheses H2 and H3 predicted a higher diversity of the idea pool when participants interacted with (H2) a question‐mode LLM and (H3) a suggestion‐mode LLM compared to (a) a model‐led LLM and (b) a vanilla LLM. As expected, participants using the question‐mode LLM showed significantly higher diversity compared to the model‐led LLM (H2a: $t(481) = 5.33$, $p < .001$, $d = 0.76$). However, no significant differences in diversity were found between participants using the question‐mode LLM and the vanilla LLM (H2b: $t(481) = 0.02$, $p = .983$, $d = 0.003$). Contrary to expectations, the suggestion-mode did not enhance idea diversity, as it showed no significant advantage over either the model-led (H3a: $t(481) = 1.87$, $p = .093$, $d = 0.29$) or the vanilla LLM (H3b: $t(481) = -3.37$, $p = .999$, $d = -0.49$). 

\begin{figure*}[t]
\Description[Outcome distributions by condition for quality, diversity, and ownership]{Three side-by-side violin plots, one each for idea quality, idea diversity, and perceived idea ownership, shown across five conditions (question-mode, suggestion-mode, model-led, vanilla, control). Each plot includes an overlaid boxplot and median marker. Idea quality is highest for the model-led and question-mode conditions relative to the others. Idea diversity is lowest for the model-led condition and higher for question-mode and vanilla. Perceived idea ownership is highest for the control condition and next highest for question-mode, with lower values for the other chatbot conditions.}
  \centering
  \includegraphics[width=\textwidth]{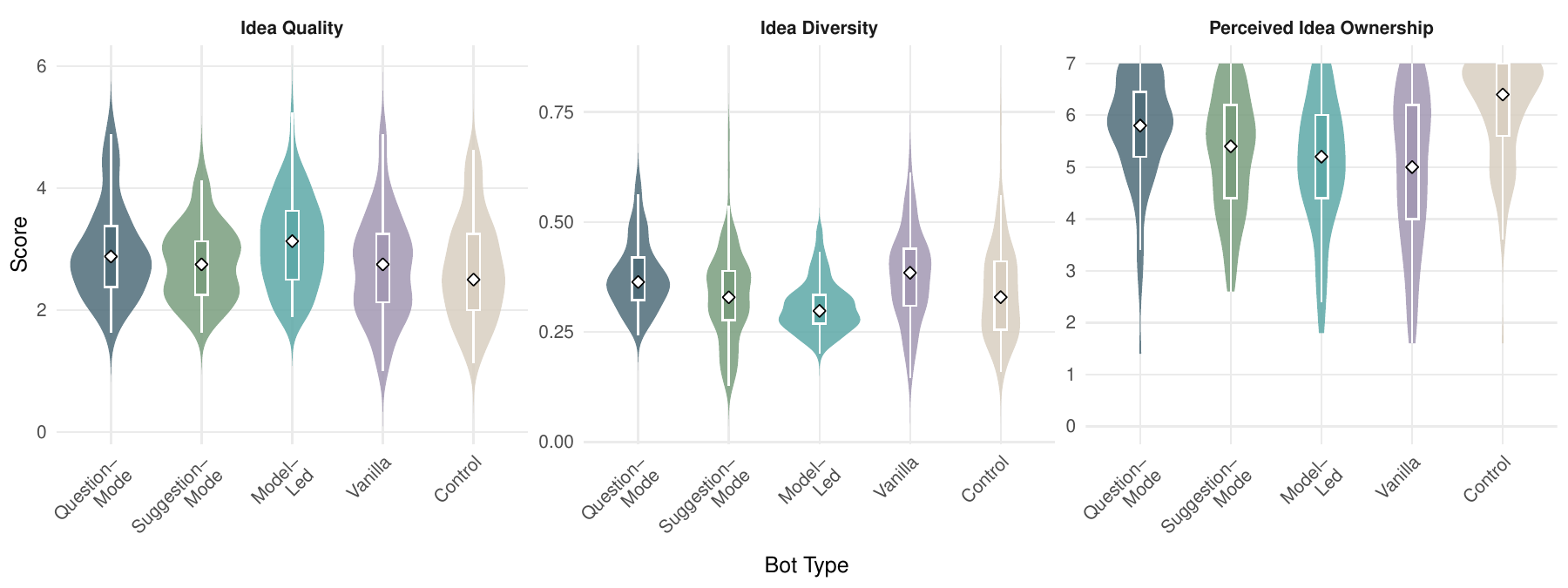}
  \caption{\textbf{Distribution of idea diversity, idea quality, and perceived idea ownership across experimental conditions.} Violin plots show the probability density of each measure, with wider sections indicating more frequent values. The boxplots show the interquartile ranges as well as the min/max, and the white dots indicate the medians.}
  \Description[Violin plots of idea diversity, idea quality, and idea ownership across conditions.]%
{The figure shows three side-by-side violin plots comparing outcomes across the five experimental conditions. The first plot displays idea diversity scores, where the vanilla LLM and the question-mode cluster around the highest median. The second plot shows idea quality, with the highest median for the question-mode and the model-led condition. The third plot shows perceived idea ownership, where the control condition and the question-mode have the highest median rating compared to the three other LLM-assisted conditions. Each violin illustrates the distribution shape, with white dots marking medians and thick black bars marking interquartile ranges.}
  \label{fig:violin-main}
\end{figure*}

The results provide \textbf{partial support} for Hypothesis H2, but \textbf{no support} for Hypothesis H3. While participants interacting with the question-mode LLM generated significantly more diverse ideas than those in the model-led condition, the suggestion-mode showed no improvement, and the vanilla LLM users already produced unexpectedly diverse ideas (Figure~\ref{fig:violin-main}, center).

\subsubsection{Perceived Idea Ownership (H4\&H5): Significantly Higher for Question-Mode Condition.}

Hypotheses H4 and H5 address perceived idea ownership when interacting with different LLM conditions  (Figure~\ref{fig:violin-main}, right). Here, we expected a higher perceived idea ownership when participants interact with a question-mode LLM and a suggestion-mode LLM compared to (a) a model-led LLM and (b) a vanilla LLM. Participants using the question‐mode LLM showed significantly higher perceived idea ownership than those interacting with a model‐led LLM (H4a: $t(481) = 3.97$, $p < .001$, $d = 0.57$) and than those using a vanilla LLM (H4b: $t(481) = 4.15$, $p < .001$, $d = 0.60$). Participants using the suggestion‐mode LLM did not differ significantly from the model‐led LLM (H5a: $t(481) = 1.22$, $p = .151$, $d = 0.18$) or from the vanilla LLM (H5b: $t(481) = 1.44$, $p = .151$, $d = 0.21$). Further, values for perceived idea ownership were relatively high for participants not interacting with any LLM. In summary, Hypothesis H4 was \textbf{fully supported}, whereas Hypothesis H5 was \textbf{not supported}. Interacting with the question-mode, but not with the suggestion-mode, significantly increased perceived ownership compared to the model-led and vanilla LLM conditions.

\begin{table*}[htbp]
\centering
\caption{\textbf{Summary of hypothesis tests}. Reported are $t$‐statistics (with degrees of freedom) and associated $p$‐values for each planned comparison. \textcolor{ForestGreen}{\ding{51}} indicates the hypothesis was supported ($p<0.05$); \textcolor{BrickRed}{\ding{55}} indicates it was not. Significance levels: $^{***}\!:\,p<0.001$; $^{**}\!:\,p<0.01$; $^{*}\!:\,p<0.05$.}
\label{tab:hypothesis_tests}
\footnotesize
\begin{tabular}{@{}llcccc@{}}
\toprule
Hyp. & Comparison & $t$ (df) & $p$ & $d$ & Supported? \\ 
\midrule
\multicolumn{6}{l}{\textit{Idea quality}}\\
H1a & \questionbadge{} Question‐mode $>$ \controlbadge{} control & $2.49\;(481)$ & $.020^{*}$ & $0.36$ & \textcolor{ForestGreen}{\ding{51}} \\
H1b & \suggestionbadge{} Suggestion‐mode $>$ \controlbadge{} control & $0.91\;(481)$ & $.366$ & $0.13$ & \textcolor{BrickRed}{\ding{55}} \\
H1c & \modelbadge{} Model‐led $>$ \controlbadge{} control & $3.86\;(481)$ & $<.001^{***}$ & $0.55$ & \textcolor{ForestGreen}{\ding{51}} \\
H1d & \vanillabadge{} Vanilla $>$ \controlbadge{} control & $0.13\;(481)$ & $.449$ & $0.02$ & \textcolor{BrickRed}{\ding{55}} \\ 
\midrule
\multicolumn{6}{l}{\textit{Idea diversity}}\\
H2a & \questionbadge{} Question‐mode $>$ \modelbadge{} Model‐led & $5.33\;(481)$ & $<.001^{***}$ & $0.76$ & \textcolor{ForestGreen}{\ding{51}} \\
H2b & \questionbadge{} Question‐mode $>$ \vanillabadge{} Vanilla & $0.02\;(481)$ & $.984$ & $0.003$ & \textcolor{BrickRed}{\ding{55}} \\
H3a & \suggestionbadge{} Suggestion‐mode $>$ \modelbadge{} Model‐led & $1.87\;(481)$ & $.093$ & $0.27$ & \textcolor{BrickRed}{\ding{55}} \\
H3b & \suggestionbadge{} Suggestion‐mode $>$ \vanillabadge{} Vanilla & $-3.37\;(481)$ & $.999$ & $-0.49$ & \textcolor{BrickRed}{\ding{55}} \\
\midrule
\multicolumn{6}{l}{\textit{Perceived idea ownership}}\\
H4a & \questionbadge{} Question‐mode $>$ \modelbadge{} Model‐led & $3.97\;(481)$ & $<.001^{***}$ & $0.57$ & \textcolor{ForestGreen}{\ding{51}} \\
H4b & \questionbadge{} Question‐mode $>$ \vanillabadge{} Vanilla & $4.15\;(481)$ & $<.001^{***}$ & $0.60$ & \textcolor{ForestGreen}{\ding{51}} \\
H5a & \suggestionbadge{} Suggestion‐mode $>$ \modelbadge{} Model‐led & $1.22\;(481)$ & $.151$ & $0.18$ & \textcolor{BrickRed}{\ding{55}} \\
H5b & \suggestionbadge{} Suggestion‐mode $>$ \vanillabadge{} Vanilla & $1.44\;(481)$ & $.151$ & $0.21$ & \textcolor{BrickRed}{\ding{55}} \\
\bottomrule
\end{tabular}
\end{table*}

\subsection{Influences of the Different Conditions on Perceived Cognitive Workload}
\begin{figure}[htbp]
 \Description[Perceived cognitive workload distributions across conditions]{Single violin plot of perceived cognitive workload on a zero to one hundred scale across five conditions (question-mode, suggestion-mode, model-led, vanilla, control), with boxplots and medians overlaid. Question-mode and suggestion-mode show higher workload distributions than model-led and vanilla. Model-led has the lowest median workload, while control and the human-led modes show higher medians and wider spreads.}
  \centering
  \includegraphics[width=0.5\textwidth]{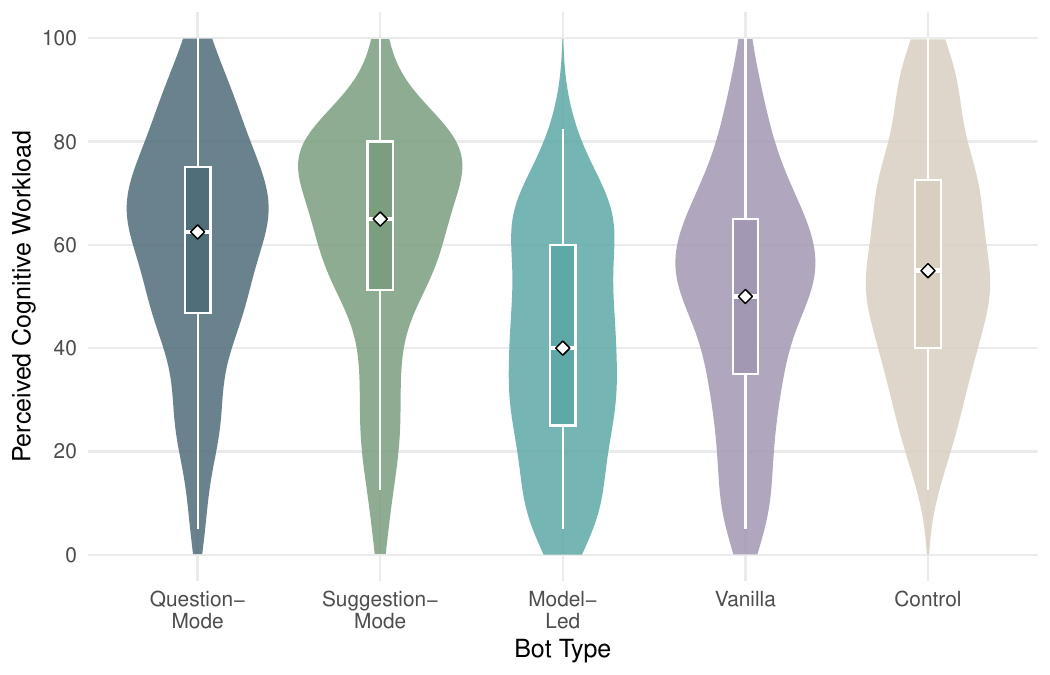}
  \caption{\textbf{Distribution of perceived cognitive workload across experimental conditions.} Violin plots show the probability density of each measure, with wider sections indicating more frequent values. The boxplots show the interquartile ranges as well as the min/max, and the white dots indicate the medians.}
  \Description[Violin plots of perceived cognitive workload across conditions.]%
{The figure shows violin plots comparing perceived cognitive workload across the five experimental conditions.  The human-led modes are higher in median workload than the control and vanilla conditions. Each violin depicts the distribution shape, white dots indicate medians, and thick black bars show interquartile ranges.}
  \label{fig:violin-cognitive}
\end{figure} 

In an exploratory analysis, we examined whether participants using the human-led LLMs perceived a higher cognitive workload than those using the model-led or vanilla LLM. To test this, we again conducted Holm-adjusted contrasts (detailed results are provided in Appendix~\ref{sec:workload}). Results showed that participants using the question-mode perceived significantly higher cognitive workload than participants using the model-led LLM ($t(481) = 6.04$, $p < .001$, $d = 0.86$)  or the vanilla LLM ($t(481) = 3.13$, $p = .004$, $d = 0.45$). A similar pattern emerged for the suggestion-mode condition, which also yielded significantly higher workload compared to both model-led ($t(481) = 5.47$, $p < .001$, $d = 0.78$)  and vanilla LLM ($t(481) = 2.58$, $p = .010$, $d = 0.37$). \new{Notably, these workload differences align with the intended interaction designs, suggesting that participants in human-led modes perceived higher cognitive workload compared to participants interacting with the model-led conditions.} We further ran a linear regression predicting perceived ownership from cognitive workload, which was a significant positive predictor ($\beta = 0.17$, $t(484) = 3.83$, $p < .001$). However, the model explained only a small proportion of the variance in perceived ownership ($R^2 = .03$). 


\subsection{Within-Subject Changes Regarding Idea Diversity}

\begin{figure*}[ht]
  \Description[Conceptual within-subject comparison from initial to refined ideas]{Two-panel illustration of within-subject change. Panel A (pre) shows a participant writing an initial idea, with a thought bubble containing a cluster of points and one highlighted point representing the initial idea. Panel B (post) shows the participant after interacting with a chatbot and producing a refined idea, with the highlighted point shifted to a new position in the thought-bubble space to indicate change from initial to refined idea.}
  \centering
  \includegraphics[width=0.9\linewidth]{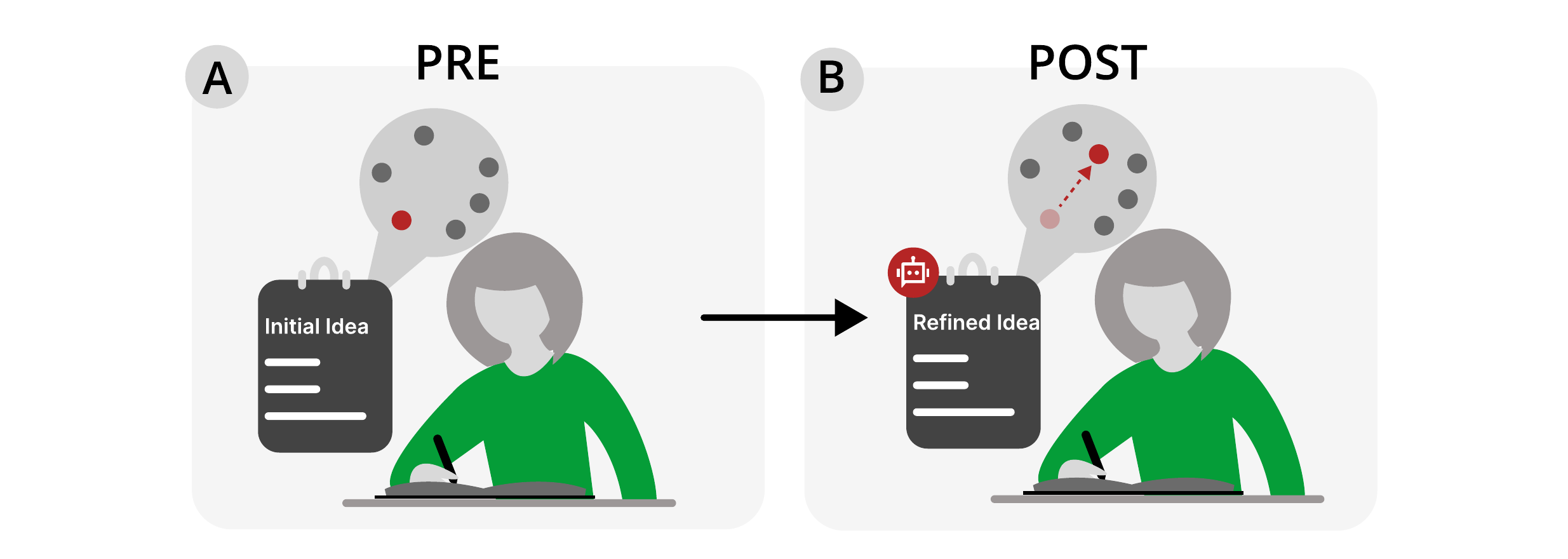}
  \Description[Illustration of a pre-post design.] {Panel A (Pre) shows a person writing down an initial idea, depicted as a single red dot among gray circles. Panel B (Post) shows the same person after interacting with an AI system, refining the idea. The refined idea is marked with a red icon, and the thought bubble illustrates the movement from the initial red dot to a more elaborated version, highlighting the cosine distance change of the ideas within the embedding space.}
  \caption{\textbf{Within-subject analysis.} \textbf{(A)} We compare the idea diversity of the initial ideas without interaction with the LLM with \textbf{(B)} the idea diversity of the refined ideas after interacting with the LLM}
  \label{fig:within-subject}
\end{figure*}

As an additional exploratory analysis, we examined within-subject changes in idea diversity across the human-led (question, suggestion) and model-led conditions (Figure~\ref{fig:within-subject}). Idea diversity was calculated for each participant’s initial idea and compared with the diversity of their refined idea, both measured relative to the centroid of all initial ideas in that condition.\footnote{Semantic similarity was calculated using cosine distance between OpenAI text-embedding-3-large vectors. Within-condition diversity measured each idea's distance to the centroid of all other ideas within the same experimental condition (leave-one-out). For the refined ideas, we compared each refined idea's embedding to the centroid of the initial ideas within the same condition.} Using the same centroid for both stages, we evaluated how interaction with the different models changed idea diversity (detailed results in Appendix \ref{sec:Within_Subject_App}). We used linear mixed-effects with planned contrasts of estimated marginal means to analyze changes in idea diversity. We found significant differences. Whereas for the model-led ($\Delta = -0.09$, 95\% CI $[-0.11, -0.07]$, $t(295) = -9.80$, $p < .001$) and suggestion-mode condition ($\Delta = -0.03$, 95\% CI $[-0.04, -0.01]$, $t(295) = -2.67$, $p = .006$), diversity decreased significantly, diversity significantly increased for the question-mode ($\Delta = 0.07$, 95\% CI $[0.05, 0.09]$, $t(295) = 7.20$, $p < .001$). In short, interaction with the question-mode expanded idea diversity, whereas both the model-led and suggestion-mode narrowed it.

\begin{figure*}[htbp]
 \Description[Trajectories of diversity scores from initial to refined ideas]{Three small line plots, one per condition (question-mode, suggestion-mode, model-led). The horizontal axis shows idea stage (initial ideas to refined ideas) and the vertical axis shows diversity score. Thin gray lines show individual participant trajectories; a thick line shows the condition mean; a dotted horizontal line shows the overall mean across conditions. The mean diversity score increases from initial to refined ideas in question-mode, decreases in suggestion-mode, and decreases more strongly in model-led.}
  \centering
  \includegraphics[width=1.0\linewidth]{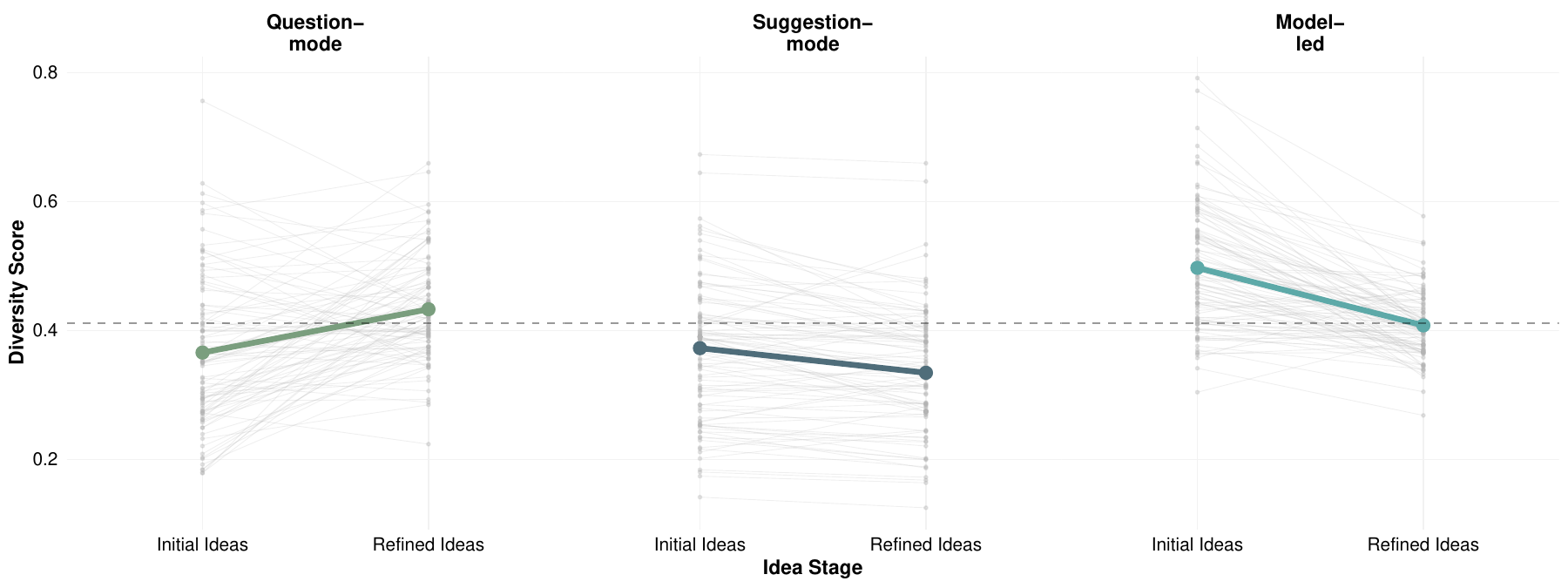}
  \caption{\textbf{Within-subject in idea diversity from initial to refined ideas.} Both initial and refined ideas are measured against the centroid of initial ideas within each condition to track within-subject movement (which we refer to as ``diversity score''). Gray lines show individual participant trajectories, thick colored lines show condition averages, and the dotted line represents the overall average across all conditions.}
  \Description{Line plot with three panels, one for each condition: Question-mode, Suggestion-mode, and Model-led. The y-axis shows diversity scores, and the x-axis shows stages from initial to refined ideas. Each panel includes many thin gray lines connecting initial and refined ideas for individual participants, and one thick colored line showing the average trend per condition. In Question-mode, the average line slopes slightly upward, showing an increase in diversity. In Suggestion-mode and Model-led, the average lines slope downward, showing a decrease in diversity. A horizontal dashed line marks the overall mean across conditions.}

  \label{fig:within_subject_change}
\end{figure*}

 \subsection{Analysis of the Interaction Mode Perceptions and Elaboration Depth}

\begin{figure*}[htbp]
  \Description[Radar chart comparing user experience, workload, ownership, and elaboration across modes]{Radar chart with seven axes: performance expectancy, elaboration depth, perceived cognitive workload, creative self-efficacy, perceived ownership, hedonic motivation, and effort expectancy. Four polygons represent question-mode, suggestion-mode, model-led, and vanilla. Question-mode is highest on elaboration depth and perceived ownership and remains high on the user experience axes. Model-led shows the lowest elaboration depth and lower ownership than the human-led modes. Vanilla and suggestion-mode fall between these profiles across most dimensions.}
  \centering
  \includegraphics[width=0.9\linewidth]{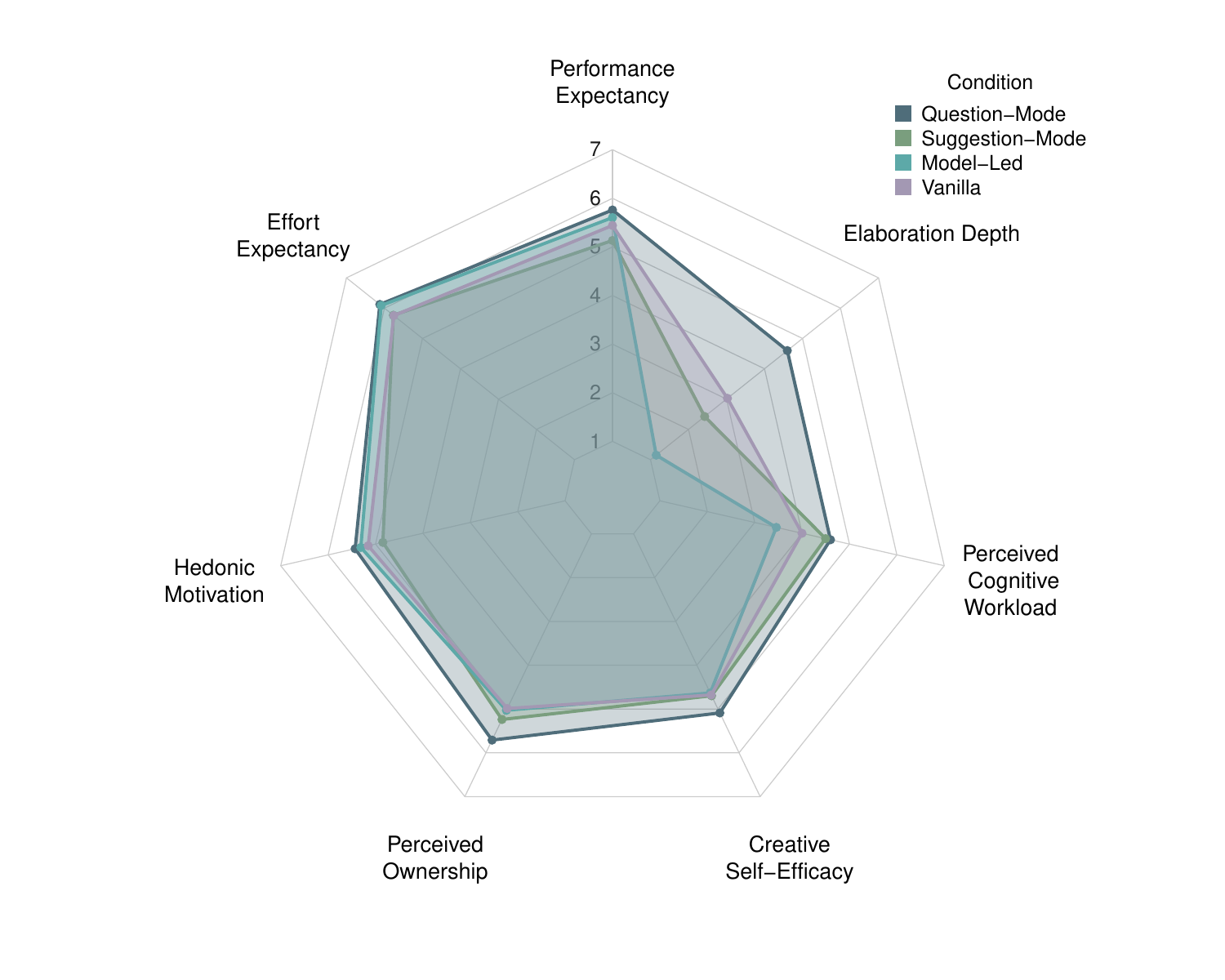}
    \Description{Radar chart with four overlapping polygons, each representing an AI interaction condition: Question-mode, Suggestion-mode, Model-led, and Vanilla. Axes include Performance Expectancy, Elaboration Depth, Mental Effort, Mental Demand, Perceived Ownership, Hedonic Motivation, and Effort Expectancy. Question-mode shows higher scores across most axes, especially elaboration depth and ownership. Suggestion-mode and Vanilla have moderate levels across all dimensions, while Model-led scores lower on elaboration depth and perceived ownership.}

  \caption{\textbf{Comparative profile of user experience and cognitive demand.} The radar chart displays mean scores across seven dimensions for each interaction mode (distinguished by color). The user experience-related measures as well as perceived ownership and creative self-efficacy were measured on 7-point Likert scales. Elaboration depth (derived from coding levels 1--5) and perceived cognitive workload (mean of mental demand and mental effort, 0--100) were linearly rescaled to a 1--7 range to facilitate visual comparison.}
  \label{fig:Spider_Chart}
\end{figure*}

To obtain further insights into the user interactions, we qualitatively coded \new{a randomly selected subset of 20 conversations per condition (80 ideas in total)} to assess the elaboration depth of participants' contributions. \new{Two trained raters independently assessed all ideas. Initial inter-rater agreement was substantial (weighted $\kappa = 0.628$). Disagreements were resolved through discussion to reach consensus, and these consensus scores were used in all subsequent analyses.} The majority of participants in the model-led condition tended to simply accept the model refinement without further elaboration (90\% of participants). Interestingly, people in the suggestion-mode condition did not deeply elaborate on the LLM suggestions but rather added some details, like additional features, to their ideas, which were already suggested by the LLM (25\% did not elaborate at all, 55\% with minor elaboration). In the question-mode, however, people substantially elaborated based on the interaction with the LLM (60\%), and 40\% even went further and added an incrementally new idea. For instance, one participant initially proposed repurposing car cameras for video calls. After the LLM pointed out its low-light and infrared capabilities, the participant expanded the idea into using such cameras for threat detection and personal security. As depicted in Figure~\ref{fig:Spider_Chart}, the question-mode successfully leveraged higher mental demand and effort to drive deeper idea elaboration. Notably, the additional cognitive workload that had to be spent on interacting with the question-mode did not diminish user experience. Instead, here we descriptively even observed higher hedonic motivation, along with greater effort and performance expectancy compared to the other conditions. In contrast, participants interacting with the suggestion-mode had high perceived cognitive workload that did not translate into high elaboration depth.
\new{To examine whether qualitative patterns linked to quantitative outcomes, we correlated elaboration depth with our dependent variables. Results showed that deeper elaboration was associated with greater idea diversity ($r = 0.28$, $p = .032$), but not with quality or ownership (both $p > .15$). Hence, individuals who have a high elaboration depth do not experience the homogenizing effect of LLM interaction, resulting in diverse ideas.} 

\section{Validation Study}
\label{sec:Validation}

\subsection{\new{Experiment Design}}

\new{
To test the robustness of our results and their generalizability to another domain, we conducted a preregistered replication study
(\url{https://aspredicted.org/de6md8.pdf}). In this study, we focus on a different task that is less stylized and, further, more relevant to management practice. Specifically, we adapted the product ideation task from \citet{girotra_ideas_2023}, asking participants to generate a product idea for a physical good that is targeted for university students in the United Kingdom, with a retail price of less than GBP~40. We selected this task to evaluate our interaction modes outside the typical AUT context and in a more realistic domain.}

\new{In this study, we restricted the initial idea to 15 words across all conditions (i.e., both model-led and human-led conditions) to improve comparability of conditions. We also implemented an additional \textbf{\emph{model-led iterative}} condition. This condition mirrored the multi-step structure used in the suggestion-mode and question-mode: after submitting their initial idea, participants were provided three suggestions based on the initial idea in the exploration stage, then they moved on to the elaboration stage, where they could choose between the first choice and two further suggestions after being able to refine the final idea before submitting.}

\new{The dependent measures were identical to those used in the initial study. In this applied setting, we additionally assessed the potential future value creation of ideas. Following prior work, we added a single intention-to-buy item to the rating instrument \cite{girotra_ideas_2023}. We also included one item assessing perceived contribution (``self vs. AI'') using a 0–100 slider scale to capture how participants attributed authorship (see results in Appendix~\ref{sec:purchaseintent}). All adapted prompts used in this validation study are provided in Appendix~\ref{sec:prompts}.}

\new{For the evaluation of idea quality and intention-to-buy, we recruited $N = 320$ raters with active student status via Prolific (each rating 20 ideas). We used students as evaluators because they are representative of the target consumers and thus well suited to also judge the perceived value of the ideas \cite{kornish_importance_2014}. Our final sample in the validation study consisted of $N = 640$ participants across six conditions.}

\begin{table*}[htbp]
\centering
\color{black}
\caption{\textbf{Summary of hypothesis tests (Study 2)}. \new{Reported are $z$-statistics for the ordinal mixed model (idea quality) and $t$-statistics with degrees of freedom for linear models (idea diversity and perceived ownership), along with associated Holm-adjusted $p$-values and Cohen's $d$ effect sizes. \textcolor{ForestGreen}{\ding{51}} indicates the hypothesis was supported ($p<0.05$); \textcolor{BrickRed}{\ding{55}} indicates it was not. Significance levels: $^{***}\!:\,p<0.001$; $^{**}\!:\,p<0.01$; $^{*}\!:\,p<0.05$.}}
\label{tab:hypothesis_tests_study2}
\footnotesize
\begin{tabular}{@{}llcccc@{}}
\toprule
Hyp. & Comparison & $z$/$t$ (df) & $p$ & $d$ & Supported? \\ 
\midrule
\multicolumn{6}{l}{\textit{Idea quality }}\\
H1a & \questionbadge{} Question-mode $>$ \controlbadge{} Control & $5.59$ & $<.001^{***}$ & $0.49$ & \textcolor{ForestGreen}{\ding{51}} \\
H1b & \suggestionbadge{} Suggestion-mode $>$ \controlbadge{} Control & $0.41$ & $.684$ & $0.04$ & \textcolor{BrickRed}{\ding{55}} \\
H1c & \modelbadge{} M-led (1-shot) $>$ \controlbadge{} Control & $6.38$ & $<.001^{***}$ & $0.50$ & \textcolor{ForestGreen}{\ding{51}} \\
H1d & \miterbadge{} M-led (iter.) $>$ \controlbadge{} Control & $4.48$ & $<.001^{***}$ & $0.39$ & \textcolor{ForestGreen}{\ding{51}} \\
H1e & \vanillabadge{} Vanilla $>$ \controlbadge{} Control & $-0.94$ & $.826$ & $-0.08$ & \textcolor{BrickRed}{\ding{55}} \\
\midrule
\multicolumn{6}{l}{\textit{Idea diversity }}\\
H2a & \questionbadge{} Question-mode $>$ \modelbadge{} M-led (1-shot) & $3.01\;(634)$ & $.006^{**}$ & $0.39$ & \textcolor{ForestGreen}{\ding{51}} \\
H2b & \questionbadge{} Question-mode $>$ \miterbadge{} M-led (iter.) & $1.01\;(634)$ & $.400$ & $0.14$ & \textcolor{BrickRed}{\ding{55}} \\
H2c & \questionbadge{} Question-mode $>$ \vanillabadge{} Vanilla & $-3.35\;(634)$ & $.999$ & $-0.48$ & \textcolor{BrickRed}{\ding{55}} \\
H3a & \suggestionbadge{} Suggestion-mode $>$ \modelbadge{} M-led (1-shot) & $8.00\;(634)$ & $<.001^{***}$ & $1.04$ & \textcolor{ForestGreen}{\ding{51}} \\
H3b & \suggestionbadge{} Suggestion-mode $>$ \miterbadge{} M-led (iter.) & $5.54\;(634)$ & $<.001^{***}$ & $0.79$ & \textcolor{ForestGreen}{\ding{51}} \\
H3c & \suggestionbadge{} Suggestion-mode $>$ \vanillabadge{} Vanilla & $1.11\;(634)$ & $.400$ & $0.16$ & \textcolor{BrickRed}{\ding{55}} \\
\midrule
\multicolumn{6}{l}{\textit{Perceived idea ownership}}\\
H4a & \questionbadge{} Question-mode $>$ \modelbadge{} M-led (1-shot) & $3.76\;(634)$ & $<.001^{***}$ & $0.49$ & \textcolor{ForestGreen}{\ding{51}} \\
H4b & \questionbadge{} Question-mode $>$ \miterbadge{} M-led (iter.) & $9.32\;(634)$ & $<.001^{***}$ & $1.33$ & \textcolor{ForestGreen}{\ding{51}} \\
H4c & \questionbadge{} Question-mode $>$ \vanillabadge{} Vanilla & $6.91\;(634)$ & $<.001^{***}$ & $1.00$ & \textcolor{ForestGreen}{\ding{51}} \\
H5a & \suggestionbadge{} Suggestion-mode $>$ \modelbadge{} M-led (1-shot) & $1.58\;(634)$ & $.058$ & $0.20$ & \textcolor{BrickRed}{\ding{55}} \\
H5b & \suggestionbadge{} Suggestion-mode $>$ \miterbadge{} M-led (iter.) & $7.36\;(634)$ & $<.001^{***}$ & $1.05$ & \textcolor{ForestGreen}{\ding{51}} \\
H5c & \suggestionbadge{} Suggestion-mode $>$ \vanillabadge{} Vanilla & $4.96\;(634)$ & $<.001^{***}$ & $0.71$ & \textcolor{ForestGreen}{\ding{51}} \\
\bottomrule
\end{tabular}
\end{table*}

\subsection{\new{Analysis}}

\new{For our validation study, we estimated linear mixed-effects models including random intercepts for both raters and ideas. This approach adjusts for systematic differences in rating severity across raters and isolates the variance attributable to the ideas themselves. As in the initial study, we evaluated our preregistered directional hypotheses at $\alpha = .05$, applied Holm-adjusted $p$-values to control the family-wise error rate, and used simple linear regressions for our analyses regarding idea diversity and perceived idea ownership.}

\subsection{\new{Results}}

\new{The validation study confirms the findings from our main study. As in the initial study, participants interacting with a question-mode (H1a: $z = 5.59$, $p < .001$, $d = 0.49$) and model-led (one-shot) interaction mode (H1c: $z = 6.38$, $p < .001$, $d = 0.50$) produced an idea with higher rated idea quality compared to participants without LLM support. Participants in the added model-led condition with an iterative approach also produced ideas with higher quality compared to participants in the control condition (H1d: $z = 4.48$, $p < .001$, $d = 0.39$), though the effect size was slightly smaller compared to the model-led one-shot condition.}

\new{For idea diversity, participants in the question-mode condition again produced ideas with higher diversity compared to participants in the model-led (one-shot) condition (H2a: $t(634) = 3.01$, $p = .006$, $d = 0.39$). In contrast to the initial experiment, participants in the suggestion-mode also produced ideas with higher diversity compared to participants in both model-led conditions (H3a: $t(634) = 8.00$, $p < .001$, $d = 1.04$ for one-shot; H3b: $t(634) = 5.54$, $p < .001$, $d = 0.79$ for iterative). Contrary to our expectations, participants interacting with the question-mode did not produce ideas with higher diversity compared to participants in the iterative model-led condition (H2b: $t(634) = 1.01$, $p = .400$, $d = 0.14$).}

\new{Regarding perceived idea ownership, participants interacting with both human-led modes showed significantly higher perceived ownership compared to those interacting with model-led and vanilla modes. Participants interacting with the question-mode LLM perceived higher ownership than those using both model-led one-shot (H4a: $t(634) = 3.76$, $p < .001$, $d = 0.49$) and iterative (H4b: $t(634) = 9.32$, $p < .001$, $d = 1.33$) approaches. Similarly, participants interacting with the suggestion-mode LLM showed higher ownership than those using the model-led iterative approach (H5b: $t(634) = 7.36$, $p < .001$, $d = 1.05$), though the difference with the model-led one-shot approach was marginally non-significant (H5a: $t(634) = 1.58$, $p = .058$, $d = 0.20$).}

\new{We further conducted an exploratory analysis examining purchase intent ratings to assess whether AI interaction modes influenced the perceived market viability of generated ideas (see Appendix~\ref{sec:purchaseintent}). Ideas created in interaction with the question-mode ($M = 2.51$, $SD = 0.58$) and model-led iterative condition ($M = 2.50$, $SD = 0.56$) showed the highest purchase intent ratings, while the suggestion-mode showed the lowest ($M = 2.20$, $SD = 0.53$), with model-led one-shot ($M = 2.24$, $SD = 0.49$), vanilla ($M = 2.30$, $SD = 0.55$), and control ($M = 2.34$, $SD = 0.62$) conditions falling in between.}

\new{In summary, the validation study successfully replicated all key findings from the main study in a different domain. Notably, in the validation study, participants interacting with both human-led modes perceived higher idea ownership compared to those using model-led approaches and participants interacting with the iterative model-led condition did not experience the same homogenization effect of ideas as in the one-shot model-led condition.}

\section{Discussion}
\label{sec:discussion}

In our research, we aimed to explore human-LLM interaction modes to augment human capabilities in creativity tasks. We focused on the comparison between human-led and model-led interaction formats and examined their influence on the idea quality, idea diversity, and perceived idea ownership. Additionally, we explored the role of perceived cognitive workload, within-subject diversity changes, and participants' elaboration depth. For this, we conducted a randomized controlled experiment with five different conditions, where participants interacted with a proactive, model-led mode or a reflective, human-led mode (in two variants, i.e., a suggestion-mode and a question-mode). We compared these interaction modes with a vanilla LLM and a control group without LLM support. Our results show that the human-led ``flipped'' collaboration format---where the LLM asks the participant questions, and the participant is the main creativity contributor---augments human capabilities regarding all three measures. In contrast, providing the participants with suggestions as inspiration does not improve creative outcomes as expected. An exploratory analysis reveals that perceived cognitive workload is higher in the human-led conditions; however, only participants in the question-mode deeply elaborate on the LLM input. Moreover, the question-mode broadens the idea space and supports participants in generating more diverse ideas, whereas the model-led and suggestion-mode conditions lead to homogenization. \new{Overall, our findings are robust across different domains, tasks, and creativity measures (see the validation study in Section~\ref{sec:Validation}).}

\subsection{Question-Mode and Model-Led Improve Idea Quality}
Across both experiments, our results show that participants interacting with the question-mode and the model-led collaboration format create ideas with significantly higher idea quality. These findings align with previous research on human-AI co-creation \cite{lee_impact_2025, kumar_human_2025, holzner_generative_2025, doshi_generative_2024, anderson_homogenization_2024}, which suggests that humans can enhance their creative performance when collaborating with LLMs. 
 However, participants interacting with the suggestion-mode and the vanilla LLM do not create higher-quality ideas, contradicting previous findings \cite{lee_empirical_2024, doshi_generative_2024}. One explanation may be task complexity. In simple creativity tasks (such as generating alternative uses for everyday objects like a rope), LLMs can be readily integrated to extend ideas. By contrast, more open-ended and complex tasks place a higher demand on participants and interaction modes. In this context, effective interaction with the vanilla LLM requires a certain prompting experience, as the copy and paste of the task is not sufficient. Also, the usefulness of suggestions may be limited, as participants often need a certain level of domain expertise to evaluate and integrate these suggestions into their ideas.

Another explanation is related to fixation effects \cite{kohn_collaborative_2011}, where users may stick to a certain idea and not develop it further. For example, the analogies in the suggestion-mode are already provided as a fully articulated example that does not let any room for ambiguity. Such explicitness can constrain human creativity by restricting the solution space \cite{qin_timing_2025}. However, in the question-mode, open-ended questions were presented in a way that allowed ambiguity. This ambiguity allows participants to explore alternative ideas and elaborate on them further \cite{gaver_ambiguity_2003}. In this way, the question-mode might be better suited to mitigate the effects of idea fixation compared to suggestions, while, at the same time, still leveraging the generative capabilities of LLMs.

\new{Notably, the exploratory analysis in the context of product innovation for UK students revealed that the higher idea quality observed in the question-mode and model-led (one-shot) conditions translated to descriptively higher purchase intent ratings from the target group. This extends previous studies conducted by \citet{girotra_ideas_2023}, thereby demonstrating that LLMs can generate commercially viable product ideas while maintaining idea diversity through appropriate interaction design.}

\subsection{Questions Broaden Idea Diversity, Suggestions Can Constrain}

Our results consistently show that participants interacting with the question-mode LLM generate significantly more diverse ideas compared to the model-led one-shot condition. Our within-subject analysis further highlights that the question-mode not only avoids homogenizing human ideas, but can even reinforce differences: refined ideas after LLM interaction were more diverse than the initial ideas. This suggests that the homogenization effect often attributed to LLMs \cite{anderson_homogenization_2024, doshi_generative_2024} is not an inherent limitation of the models, but rather an effect of interaction design. Elaborate collaborative frameworks can thus mitigate homogenization and actively foster divergent thinking. 

\new{By contrast, the effectiveness of the suggestion-mode in supporting idea diversity might depend on the domain. Whereas in the initial AUT experiment, the suggestion-mode rather constrained the idea space, it kept idea diversity high in the validation study. A reason for that might be, that the AUT task in the first study did not inspire deeper elaboration in the suggestion-mode. Similar to \citet{kumar_human_2025}, participants might have simply adopted LLM suggestions. 
In contrast, creating product ideas requires participants to identify genuine needs and problems, enabling the suggestions as inspiration rather restriction, thereby fostering idea diversity. Our findings extend current literature on the homogenization effect by showing that not only the timing of LLM support \cite{qin_timing_2025} or prompting matters \cite{meincke_prompting_2024}, but also the specific interaction mode and the task context.}

Notably, contrary to \citet{meincke_chatgpt_2025}, even the vanilla LLM without any guiding system prompt produced relatively diverse ideas. This might be partly explained by participants generating ideas that did not strictly adhere to task constraints. Additionally, this indicates that removing strong guidance may broaden the exploration space and counter homogenization. However, the absence of scaffolding also limits opportunities for structured improvement and higher idea quality, suggesting a trade-off between diversity and guidance. \new{Interestingly, participants in the iterative model-led condition of the validation study did not exhibit homogenization, underscoring the value of iterative processes in creative work \cite{wallas_art_1926, guilford_nature_1967, frich_mapping_2019}.}

In summary, these results are notable because many existing creativity support tools---both in research and practice---are designed around execution-oriented interactions (similar to our one-shot model-led condition) or suggestion-based scaffolds \cite{doshi_generative_2024, qin_timing_2025, shaer_ai-augmented_2024}. For example, Figma's ideation \emph{Jambot}\footnote{\url{https://www.figma.com/community/widget/1274481464484630971}}, or Microsoft's \emph{Copilot} in Word\footnote{\url{https://support.microsoft.com/en-us/office/welcome-to-copilot-in-word-2135e85f-a467-463b-b2f0-c51a46d625d1}}, where ideation is scaffolded through suggested content or user instructions are executed in draft form. These tools appear, in particular, to carry a risk of homogenization. Our findings thus support recent claims of the relevance of evaluating creativity support tools via homogenization analysis \cite{anderson_evaluating_2024}. 

\subsection{Question-Mode Keeps Perceived Idea Ownership High
}
Across both experiments, our results show that participants interacting with the question-mode perceive significantly higher idea ownership compared to the model-led mode and the vanilla LLM. In line with prior research, high human involvement in the process can increase the perceived idea ownership of participants \cite{xu_what_2024}. This represents a possibility to effectively co-create with LLMs without diminishing users' perceived ownership, by keeping them in the lead of the creative work and providing support from the LLM. Thereby, we extend earlier work by showing that not only longer prompts \cite{joshi_writing_2025} or user-initiated ideation \cite{qin_timing_2025} but also question-driven interactions can preserve ownership while benefiting from LLM augmentation. \new{Interestingly, participants in the suggestion-mode only experience higher ownership in the second study in the domain of  product ideation, but not in the first study using an AUT task, mirroring the pattern observed for idea diversity.}

In summary, in the human-led conditions—particularly in the question-mode—participants were effectively supported by the LLM while maintaining high perceived idea ownership. Our findings show that, to augment human capabilities with LLMs and mitigate negative effects like reduced idea diversity and idea ownership, it is effective to place the human in the creative lead. When designing LLMs as a thought partner, ideas can emerge through the interaction, augmenting human creativity rather than replacing it.

\subsection{Theoretical Implications}

Our research advances the understanding of human–LLM co-creation by demonstrating that the way humans are involved substantially shapes interaction outcomes. Especially in the context of complex tasks, providing the task to an LLM is not effective, and more collaborative frameworks are necessary.

One central implication for HCI and LLM research is the need to move beyond “vanilla” chatbot setups or one-shot prompting. Our findings suggest how more elaborate, sustained interaction modes can advance the theoretical understanding of LLM-supported co-creation. Such richer modes of collaborative engagement allow us to balance ecological validity and experimental control. By testing more elaborate interaction modes with complex creativity tasks, we show how our HCI designers can capture the complexity of real-world creative processes while preserving the reliability and replicability of results. 

Our results also highlight that creativity unfolds across multiple levels---at the individual level (idea quality), the collective level (idea diversity), and the perception level (perceived idea ownership). Each dimension responds differently to the specific collaboration format, underscoring the importance of evaluating the effectiveness of creativity support tools on multiple dimensions.

Finally, our results could suggest that perceived cognitive workload and elaboration depth play a crucial role in co-creation: namely, that humans really contribute to the creative outcome, engage in the process, and try to enhance the results over the LLM capabilities instead of letting the AI automate. In contrast, in model-led interaction modes, humans might offload cognitive work to the AI, thereby reducing their contribution to the creative outcome.

\subsection{Practical Implications for the Design of Co-Creation Frameworks}

Although LLMs are often described as thought partners with an emphasis on co-creation as a design pattern \cite{rafner_creativity_2023, collins_building_2024}, actual design for LLM-based co-creation remains scarce. Our study provides designers of co-creation tools with three concrete design principles on how to effectively create co-creation interactions with LLMs. Based on our findings, we suggest that designers of co-creation tools should:

\begin{itemize}
    \item \textbf{Principle 1: Maintain human exploration of the idea space:} 
    \new{To integrate LLMs in co-creation applications, HCI designers should amplify human cognitive exploration, allowing users to navigate broad possibility spaces with creativity and autonomy. Our findings imply that, in creative work, a model-led interaction approach may accelerate creative processes but at the risk of diversity and perceived idea ownership. This is particularly salient during exploration stages, where human creativity offers unique strengths in navigating broad possibility spaces. In practice, this means that the human initially generates ideas, while the LLM primarily asks questions that prompt elaboration and reflection. This helps establish a mutual understanding between human and AI regarding the intentions, a key principle for effective co-creation \cite{collins_building_2024}.}
    
    \item \textbf{Principle 2: Counter cognitive offloading:}
    \new{The most effective LLM-based co-creation tools will not necessarily rely on the most capable technology, but on designs that actively support human thinking. Co-creation with model-led LLM interaction modes can lead to overreliance and cognitive offloading. This is not only important for the exploration stage, but also for evaluating the best outcome, as only an individual engaging with the output can evaluate it properly \cite{tankelevitch2024metacognitive, gmeiner_exploring_2025}. Our results suggest that designing for creativity thus requires a human-centered approach that engages humans in deep elaboration. In practice, designers should therefore aim to promote active human contribution rather than passive generation of LLM output.}
    
    \item \textbf{Principle 3: Guide users through iterative refinement with LLMs:} \new{Building on well-established theories of iterative creativity \cite{resnick_design_2005}, LLM systems for co-creation should not treat generation as a one-shot process but as an evolving collaboration between user and model. While LLMs can rapidly generate output in a single step, this approach often fails to produce optimal and diverse results. Moving forward, LLM-based co-creation tools should intentionally scaffold iterative refinement---whether led by the user or by the model---to help users explore alternatives, critique drafts, and converge on stronger solutions. Designing for this iterative loop implies that interfaces should support revision pathways (e.g., models could learn to propose structured next steps), and systems should encourage reflection and re-generation. Embedding iteration as a core interaction pattern will unlock greater creativity, higher-quality ideas, and overall more effective human–AI co-creation.}
\end{itemize}

    \new{These principles can be, for example, applied to LLM-based writing assistants for creative writing (e.g., for essays or short stories) as Microsoft's Copilot in Word. The LLM assistant can first support idea generation through a brief question-only stage that prompts users to articulate claims, audiences, and constraints, keeping early exploration human-led. During elaboration, the assistant can encourage users to outline key points before any drafting occurs, reducing cognitive offloading and preserving meaningful human contribution. During refinement, initiative might be shared between human and LLM, with the human shaping the text and the AI providing feedback and additional ideas across iterative cycles.}

\subsection{Limitations and Potential for Future Work}

One limitation of our study is that we tested our interaction modes only in text-based tasks, leaving open the question of whether these findings generalize to other, multimodal applications, such as visual design. Still, text-based ideation remains highly relevant, as most creative processes involve it to some extent, ranging from drafting lyrics to developing film plots. In addition, our participants were drawn from the general public, and it remains unclear how other groups, such as domain experts, might benefit from the different interaction modes. Additionally, for our controlled experiment, the conditions differed mainly in prompting style. However, participants could not flexibly switch or combine modes as they might in real-world use. It might be, for example, interesting to let the participant fully explore the idea based on the question-mode and later further provide suggestions. Future work could therefore explore hybrid strategies. Finally, we cannot make any conclusions about the longitudinal effects of the interaction modes (e.g., how the interaction modes may influence participants' creative processes over the course of days or weeks).

\new{Our study examined text-based divergent thinking tasks with general population participants in controlled conditions. While this design enabled rigorous hypothesis testing, this points to concrete opportunities for extending our work. Nevertheless, we validated our findings in a more realistic ideation task (see Section~\ref{sec:Validation}). Future research should build on our findings by conducting field studies in authentic work settings, testing interaction modes with domain experts, and examining how different task contexts moderate the effectiveness of human-led versus model-led approaches.} \new{First, field studies would strengthen external validity. Deploying these interaction modes in real-world creative work---such as marketing professionals developing campaigns---would test whether our laboratory findings generalize to authentic practice where creativity must balance strategic constraints (brand consistency, search engine advertising requirements, character limits). Such studies would also enable measuring real-world success metrics like conversion rates beyond creative quality ratings.} \new{Second, testing with domain experts is essential. Research suggests that experts might not only be better at generating creative ideas but also at evaluating the novelty and relevance of AI-generated outputs \cite{eisenmann_expertise_2025,jia_when_2024}, potentially affecting the effectiveness of different interaction modes. Understanding whether expertise moderates interaction mode effectiveness would clarify the boundary conditions of our findings.} \new{Finally, different task domains have different requirements. While question-mode interactions support divergent thinking, they might hinder performance in programming or problem-solving where productivity and correctness outweigh novelty. Future studies should examine interaction modes across domains and explore approaches that adaptively support human strengths \cite{collins_building_2024}.}

Beyond these specific directions, our results highlight the augmenting potential of question-mode interaction in creativity contexts. Future studies should test whether this interaction mode can also mitigate risks of reduced critical thinking \cite{lee_impact_2025} or diminished skill acquisition \cite{kumar_human_2025}. A ``flipped'' collaboration format---in which humans remain in the lead while the LLM supports reflection---may help ensure that skills persist beyond individual interactions. This raises the broader question of whether LLMs should always do everything they are capable of. In the context of co-creation, the most effective LLM is the one that best supports the user in their creativity. Yet, many applications continue to struggle with implementing LLMs in ways that deliver meaningful benefits. While more advanced models may improve outcomes, in co-creation, they constitute only one side of the equation. The other side---the design of interaction modes---still bears substantial, and largely untapped, potential.

\subsection{Concluding Remarks}
We set out to examine the effect of different interaction modes on human-LLM co-creation. Our findings show that the question-mode not only improves idea quality but also preserves perceived ownership and diversity---three central dimensions of creativity when collaborating with LLMs. These findings highlight the importance of interaction design for the capabilities of humans and LLMs. For designers, our results suggest that creativity support tools should emphasize human-led, collaborative interaction formats that stimulate reflection instead of replacing human contribution. Taken together, we position the question-mode as a promising design pattern to augment humans when collaborating with LLMs.


\begin{acks}
We would like to thank Julia Falinska for her support in designing the graphics.
\end{acks}

\section*{Data availability} The questionnaire, analysis code, and anonymized data are publicly available in a GitHub repository: \url{https://github.com/SM2982/replication-creativity-CHI2026}.

\bibliographystyle{ACM-Reference-Format}
\bibliography{Citations}
\newpage

\onecolumn
\appendix

\section{Prompts Used in the Different Conditions}
\label{sec:prompts}

\begin{table*}[htbp]
  \caption{Experimental Conditions and AI Assistant Prompts for the Initial Study}
  \label{tab:conditions_prompts}

  \begin{minipage}{\textwidth}
    \centering
    \scriptsize

\begin{tabularx}{\textwidth}{>{\raggedright\arraybackslash}p{2.2cm} >{\raggedright\arraybackslash}X}
\toprule
\textbf{Condition} & \textbf{Prompt description and key instructions} \\
\midrule

\textbf{Question-mode} & 
\textbf{Exploration Phase:} ``\textit{<Your task:>}

You act as a creative thought partner. The user has shared an idea and your goal is to help the user explore unexpected and novel ideas for the task [car repurposing]. Ask the user a thought-provoking question, to generate alternative ideas to the initial idea.

\textit{<The questions you ask>}

1. For the first alternative idea: Using the knowledge approach:

To help them reimagine or enhance this idea, briefly share 1–2 sentences of specific knowledge about either: The capabilities or lesser-known functions of the feature(s) mentioned in the idea that could make the idea even more creative or another vehicle feature that could contribute to or transform the idea. Make clear that this knowledge could inspire a new or improved version of the idea. Then ask: 'How could this insight inspire a new version or twist on your idea?'

2. For the second alternative idea: Using the analogy approach:

Shift to a more lateral thinking approach. Provide a short analogy—this could be a product, brand, use case, or concept from another domain—that resembles the core idea or solves a similar problem. Use 1–2 sentences to describe the analogy clearly. Then ask: 'How could this kind of repurposing inspire a new version or twist on your idea?'

\textit{<How you act>}

Shortly acknowledge the user input. Ask thought-provoking questions using the knowledge approach or the analogy approach. Sound like smooth conversation with a friend. Max 40 words. Only ask one question per message.'' \\
\cmidrule(r){2-2}
& \textbf{Choice Phase:} ``\textit{<Your task:>} Present all available options verbatim without further elaboration (initial idea + alternatives). Ask the user 'Please choose the most unique and creative option to develop further'. When the user makes their choice, call the choose\_alternative function.'' \\
\cmidrule(r){2-2}
& \textbf{Elaboration Phase:} ``\textit{<Your task:>} You are an expert in facilitating practical idea development. Your goal is to help them further elaborate and clarify the idea—especially when it is still abstract or vague—so it becomes more useful and actionable. Ask one focused, thought-provoking question that helps the user: Ask a question that might further increase the creativity and novelty of the idea, Clarify what problem the idea solves, Specify the context of use or the intended users, Make abstract aspects of the idea more concrete, Consider user needs, behaviors, or usage scenarios, Shift perspective to inspire elaboration, Identify what may be missing in the current idea and ask a targeted question to address this gap.'' \\
\midrule

\textbf{Model-led} & 
\textbf{System Prompt:} ``Your task is to take the user's input and write an idea by the following criteria: (1) It has to fit the task [car repurposing] and the car repurposing context (2) Increase the novelty and creativity of the idea (3) Increase the usefulness of the idea (4) Make it more elaborate (5) Provide one coherent idea. Keep your response focused on the improved idea - don't add explanations, disclaimers, or ask follow-up questions. Please keep the style and the tone of the original idea and keep it as concise as possible, maximum 40 words.'' \\
\midrule

\textbf{Suggestion-mode} & 
\textbf{Exploration Phase:} ``You are an expert in facilitating creative idea development through suggestions. The user has shared an idea for the task [car repurposing] and your goal is to help them explore it more deeply and make it more novel. Provide 3 specific suggestions for creative ideas (40-60 words each) that are similar to the user's idea but think it further and make it more novel and surprising. Your suggestions should focus on: Combining the idea with unrelated contexts, Increasing uniqueness and originality, Creating unexpected elements, Challenging assumptions, Reversing or inverting core concepts, use first principles thinking and lateral thinking, Prioritize novelty, unexpected connections and contrasting perspectives over safe ideas. Format: *Some ideas to make your concept more creative and novel:* **Suggestion 1:** [novel suggestion] **Suggestion 2:** [novel suggestion] **Suggestion 3:** [novel suggestion].'' \\
\cmidrule(r){2-2}
& \textbf{Elaboration Phase:} ``You are an expert in facilitating practical idea development through suggestions. The user has shared a creative idea for the task [car repurposing] and your goal is to help them make it more useful and elaborate. Provide 3 specific suggestions (40-60 words each) that enhance the user's idea to make it more useful. Your suggestions should focus on: Solving specific problems or addressing frustrations, Identifying clear target users and benefits, Enhancing the value proposition, Addressing implementation challenges, Creating competitive advantages, Improving commercial potential and monetization, The three suggestions should be very different from each other. Format: Some ideas to make your concept more useful and practical: **Suggestion 1:** [usefulness-focused suggestion] **Suggestion 2:** [usefulness-focused suggestion] **Suggestion 3:** [usefulness-focused suggestion].'' \\
\midrule

\textbf{Control} & 
No AI assistance provided. Participants work independently on the creativity task without any chatbot interaction. \\
\midrule

\textbf{Vanilla} & 
No specific system prompt. Standard GPT-4 model with default behavior, temperature=0.7, max 500 tokens per response. \\

\bottomrule
\end{tabularx}

    \bigskip
    \raggedright\footnotesize
    \emph{Note:} All AI conditions used GPT-4.1 with temperature=0.7.
    Question-mode and Suggestion-mode utilized tool calls to transition
    between different interaction stages, enabling controlled user flow
    and optimal stage management throughout the creative process.
  \end{minipage}
\end{table*}

\newpage

\begin{table*}[htbp]
  \caption{Experimental Conditions and AI Assistant Prompts for the Validation Study}
  \label{tab:conditions_prompts_study2}

  \begin{minipage}{\textwidth}
    \centering
    \scriptsize


\begin{tabularx}{\textwidth}{>{\raggedright\arraybackslash}p{2.2cm} >{\raggedright\arraybackslash}X}

\toprule

\textbf{Condition} & \textbf{Prompt description and key instructions} \\

\midrule

\textbf{Feedback (Question-mode)} & 

\textbf{Exploration Phase:} ``\textit{<Your task:>} You act as a creative thought partner. The user has shared an idea and your goal is to help the user explore unexpected and novel ideas for the task [product ideas for UK university students]. Ask the user a thought-provoking question, to generate alternative ideas to the initial idea.

\textit{<The questions you ask>}

1. For the first alternative idea: Using the knowledge approach: To help them reimagine or enhance this idea, briefly share 1–2 sentences of specific knowledge about: Specific product features, other products or market trends that could contribute to or transform the product of the initial idea. Make clear that this knowledge could inspire a new or improved version of the idea. Then ask: 'How could this insight inspire a new version or twist on your idea? Please propose an alternative idea.'

2. For the second alternative idea: Using the analogy approach: Shift to a more lateral thinking approach. Provide a short analogy—this could be a product, brand, use case, or concept unrelated to the initial idea, but that could inspire a new version or twist on the initial idea. Use 1–2 sentences to describe the analogy clearly. Then ask: 'How could this analogy inspire a new version or twist on your idea? Please propose an alternative idea.'

\textit{<How you act>}

Shortly acknowledge the user input. Ask thought-provoking questions using the knowledge approach or the analogy approach. Sound like smooth conversation with a friend. Max 40 words. Only ask one question per message. Write relevant words in **bold** using markdown formatting.'' \\

\cmidrule(r){2-2}

& \textbf{Choice Phase:} ``\textit{<Your task:>} Present the alternatives clearly to the user and help them choose which one to develop further. Present all available options verbatim without further elaboration (initial idea + alternatives). Write relevant words in **bold** using markdown formatting. Ask the user 'Please choose the most **unique and creative** option to develop further'. When the user makes their choice, call the choose\_alternative function.'' \\

\cmidrule(r){2-2}

& \textbf{Elaboration Phase:} ``\textit{<Your task:>} You are an expert in facilitating practical idea development. The user has shared a creative product idea for university students in the UK. Your goal is to help them further elaborate and clarify the idea—especially when it is still abstract or vague—so it becomes more useful and actionable. Briefly acknowledge the user's previous message. Ask one focused, thought-provoking question that helps the user: Ask a question that might further increase the creativity and novelty of the idea, Clarify what problem the idea solves for UK university students, Specify the context of use or the intended users, Make abstract aspects of the idea more concrete, Consider the specific needs, behaviors, habits, or preferences of UK university students and their usage scenarios, Shift perspective (e.g., different user groups, environments) to inspire elaboration, Identify what may be missing in the current idea (e.g., unclear value, user role, context) and ask a targeted question to address this gap. If the idea is vague or confusing, say so and ask a specific question to clarify it. Be critical: if something is hard to understand or doesn't seem feasible, say that clearly. Use markdown formatting **bold** to highlight key concepts or terms from the user's idea.'' \\

\midrule

\textbf{Suggestion-mode} & 

\textbf{Exploration Phase:} ``You are an expert in facilitating creative idea development through suggestions. The user has shared an idea for the task [product ideas for UK university students] and your goal is to help them explore it more deeply and make it more novel. Provide 3 specific suggestions for creative ideas (40-60 words each) that are similar to the user's idea but think it further and make it more novel and surprising. Your suggestions should focus on: Combining the idea with unrelated contexts, Increasing uniqueness and originality, Creating unexpected elements, Challenging assumptions, Reversing or inverting core concepts, use first principles thinking and lateral thinking, Prioritize novelty, unexpected connections and contrasting perspectives over safe ideas. Format: *Some ideas to make your concept more creative and novel:* **Suggestion 1:** [novel suggestion] **Suggestion 2:** [novel suggestion] **Suggestion 3:** [novel suggestion].'' \\

\cmidrule(r){2-2}

& \textbf{Elaboration Phase:} ``You are an expert in facilitating practical idea development through suggestions. The user has shared a creative idea for the task [product ideas for UK university students] and your goal is to help them make it more useful and elaborate. Provide 3 specific suggestions (40-60 words each) that enhance the user's idea to make it more useful. Your suggestions should focus on: Solving specific problems or addressing frustrations of UK university students, Identifying clear target users and benefits of UK university students, Enhancing the value proposition, Addressing implementation challenges of UK university students, Creating competitive advantages, Improving commercial potential and monetization, The three suggestions should be very different from each other. Format: Some ideas to make your concept more useful and practical: **Suggestion 1:** [usefulness-focused suggestion] **Suggestion 2:** [usefulness-focused suggestion] **Suggestion 3:** [usefulness-focused suggestion].'' \\

\midrule

\textbf{Model-led (one-shot)} & 

\textbf{System Prompt:} ``\textit{<Your task>} Your task is to take the user's input and write an idea by the following criteria: (1) It has to fit the task [product ideas for UK university students] and the product idea context for university students in the UK (2) Increase the novelty and creativity of the idea (3) Increase the usefulness of the idea (4) Make it more elaborate (5) Provide one coherent idea. \textit{<Your Response>} Keep your response focused on the improved idea - don't add explanations, disclaimers, or ask follow-up questions. Please keep the style and the tone of the original idea and keep it as concise as possible, maximum 40 words.'' \\

\midrule

\textbf{Model-led (iterative)} & 

\textbf{Exploration Phase:} ``You are an expert in facilitating creative idea development through suggestions. The user has shared an idea for the task [product ideas for UK university students] and your goal is to help them explore it more deeply and make it more novel. Provide 3 specific alternative creative ideas (40-60 words each) that are inspired by the user's idea but think it further and make it more novel and surprising. Your suggestions should focus on: Combining the idea with unrelated contexts, Increasing uniqueness and originality, Creating unexpected elements, Challenging assumptions, Reversing or inverting core concepts, use first principles thinking and lateral thinking, Prioritize novelty, unexpected connections and contrasting perspectives over safe ideas, Each suggestion must be self-contained and readable as a standalone proposition. Format: *Some ideas to make your concept more creative and novel:* **Suggestion 1:** [novel suggestion] **Suggestion 2:** [novel suggestion] **Suggestion 3:** [novel suggestion]. After the user selects and refines one of these suggestions, automatically move to the next stage.'' \\

\cmidrule(r){2-2}

& \textbf{Elaboration Phase:} ``You are an expert in facilitating practical idea development through suggestions. The user has shared a creative idea for the task [product ideas for UK university students] and your goal is to help them make it more useful and elaborate. Provide 2 specific suggestions for creative ideas (40-60 words each) that are similar to the user's idea but think it further and make it more novel and surprising. Your suggestions should focus on: Solving specific problems or addressing frustrations of UK university students, Identifying clear target users and benefits of UK university students, Enhancing the value proposition, Addressing implementation challenges of UK university students, Creating competitive advantages, Improving commercial potential and monetization, The two suggestions should be very different from each other, Each suggestion must be self-contained and readable as a standalone proposition. Format: Some ideas to make your concept more useful and practical: **Suggestion 1 (Your Refined Idea):** [user's refined idea] **Suggestion 2:** [usefulness-focused suggestion] **Suggestion 3:** [usefulness-focused suggestion].'' \\

\bottomrule

\end{tabularx}

    \bigskip
    \raggedright\footnotesize
    \emph{Note:} All AI conditions used GPT-4.1 with temperature=0.7.
    Question-mode and Suggestion-mode utilized tool calls to transition
    between different interaction stages, enabling controlled user flow
    and optimal stage management throughout the creative process.
  \end{minipage}
\end{table*}

\newpage
\section{Examples of Ideas (Initial Study)}

\begin{table*}[htbp]
  \caption{Examples of refined ideas by condition (three per condition)}
  \label{tab:refined_ideas_examples}
  \centering
  \scriptsize
\scriptsize
\begin{tabularx}{\textwidth}{>{\raggedright\arraybackslash}p{1.7cm} >{\centering\arraybackslash}p{0.6cm} >{\raggedright\arraybackslash}X}
  \toprule
  \textbf{Condition} & \textbf{\#} & \textbf{Refined idea example} \\
  \midrule

  \multirow{3}{*}{Question-mode}
    & 1 & Car lights with selectable messages, operated from the steering wheel, for use in city streets and traffic congestion, allowing drivers to communicate issues or intentions that can't be shown with regular lights—such as 'bike ahead,' 'sorry,' or humorous icons—to reduce misunderstandings, increase safety, and add a fun, appealing element to driving in stressful or busy environments. \\
    \cmidrule(lr){2-3}
    & 2 & Solar panel wipers use image recognition technology from car cameras to automatically detect when cleaning is needed and activate accordingly, providing a sophisticated and efficient self-maintenance system for all users of solar panels, including residential, commercial, and industrial settings. \\
    \cmidrule(lr){2-3}
    & 3 & Screens in cars display real-time information about the driver's current location to enhance driver awareness (e.g., warning of hazards or recent incidents nearby), sensors detect life movement such as animals or children to prevent hit and runs, and lights are used to communicate intentions between drivers using specific signals. These features are especially useful for driving at night and in school zones. \\
  \midrule

  \multirow{3}{*}{model-led}
    & 1 & Transform decommissioned cars into covert mobile surveillance units equipped with hidden cameras, audio sensors, and license plate readers, allowing police to discreetly monitor high-risk areas and gather real-time evidence against drug trafficking operations. \\
    \cmidrule(lr){2-3}
    & 2 & Install a network of former car reversing sensors along the home's facade, integrating them with smart home systems to detect approaching individuals or objects, triggering customized alerts, lighting adjustments, or automated security responses for enhanced safety and automation. \\
    \cmidrule(lr){2-3}
    & 3 & Transform the car interior into a modular living space by engineering seats that not only recline flat but also interlock, creating a seamless surface that can be converted into a bed, workspace, or lounge area as needed. \\
  \midrule

  \multirow{3}{*}{Suggestion-mode}
    & 1 & Cameras to pickup animals and other dangerous things / warnings and potentially give a warning or an automatic stopping feature / system. Solar pannels to charge anything related to the car, cameras, phones etc which can be useful, Automated brakes for stop lights as well which have many sensors around the car. Automatic system to tell you where the next refuel is, stations and services and also triggers environmental warnings, too cold, maybe what to look out for, any weather warnings \\
    \cmidrule(lr){2-3}
    & 2 & removable seat covers could be repurposed to use as a hammaock for a small child (while car is parked) or as a sling to use for storage like they have in camper vans, they could have even more uses if made of waterroof fabric (at least on one side) to use as an emrgency blanket or poncho \\
    \cmidrule(lr){2-3}
    & 3 & Car proximity sensors could be repurposed into an indoor fall detection and prevention system for elderly care. Installed discreetly around a home, these sensors would monitor movement patterns and detect sudden shifts in motion or unusual inactivity. If someone falls or stops moving unexpectedly, the system could alert caregivers instantly. Unlike traditional cameras, proximity sensors are less intrusive, preserving privacy while maintaining safety. Additionally, these sensors could be integrated with floor-level lighting to illuminate pathways when movement is detected at night, reducing trip hazards and improving independence for elderly individuals. To increase accuracy and reduce false alarms, the system could be paired with wearable devices—such as smart bracelets or pendants—that track heart rate, motion, or body temperature. Cross-referencing this biometric data with sensor feedback would enable earlier detection of potential health issues and deliver proactive alerts to caregivers. A modular, wireless kit could be designed for easy, tool-free installation in rental properties or older homes, making it accessible to a broader aging-in-place population. Beyond safety, the system could offer adaptive "comfort zones," adjusting lighting, sound, or visuals to create soothing environments that respond to presence—promoting wellness for both elderly users and the wider household. With customizable modes for different age groups, the platform could evolve into a full-featured, privacy-first smart wellness system for multi-generational homes. \\
  \midrule

  \multirow{3}{*}{Control}
    & 1 & Many cars have some form of decorative lighting inside. My idea would be to utilise this as part of the hazard warning process. For example - the car senses a hazard ahead and turns a subtle shade of yellow or red depending on the hazard. Or when the car is braking.. the lights turn a shade of yellow to indicate the car is slowing and red at a full stop. The car could glow a subtle shade of green to indicate the car is in good working order or perhaps orange to indicate a fault. Just subtle colour changes so the user is intune with the car. A more interactive driving experience. \\
    \cmidrule(lr){2-3}
    & 2 & cameras - make smart (connect to internet) and use to capture wildlife that visits the car (motion activated). Could also use camera to provide images for geolocation, thus double-checking where the car is. Speakers could be used in combination with the car's phone microphone to detect and amplify sounds, so any woldlife could be identified, and sounds played to either deter it or attract it, depending on the user's desire. \\
    \cmidrule(lr){2-3}
    & 3 & Sensors could help alert you if something spills in the car. Speakers could tell your children to calm down or ask them matches questions to keep them entertained Lights could be used on the ceiling of the car to help the toddler go to sleep \\
  \midrule

  \multirow{3}{*}{Vanilla}
    & 1 & Repurpose old car seats to make comfortable seating area outside, adding some height to allow for those unable to get down low. Using car headlights as a great way to add light to a garden! Old dash cams to be turned in to a wildlife camera to allow you to see the wildlife as it happens! I would use a car steering wheel to create a garden feature! Dash cams could also be used indoors to monitor movement around the home for example keeping an eye on elderly parents/grandparents! \\
    \cmidrule(lr){2-3}
    & 2 & Audio Book Club: Use the Bluetooth or CD player to host a “car book club”—listen to chapters with friends and discuss, even if you’re parked. Could link this via WIFI and have a joint book club reading. Kinda like an open air cinema but with books \\
    \cmidrule(lr){2-3}
    & 3 & We could re-purpose cameras and sensors for wildlife capture cameras. So they can be installed in a public garden to photograph wild animals at night. Lights and speakers could also be used to make it a sensory garden during the day for humans. \\

  \bottomrule
\end{tabularx}
\end{table*}

\newpage
\section{Correlations (Initial Study)}

\begin{table*}[htbp]
  \caption{Means, standard deviations, and  of the main study}
  \label{tab:correlations}

  \begin{minipage}{\textwidth}
    \centering
    \scriptsize

    \scriptsize
\setlength{\tabcolsep}{3pt}
\renewcommand{\arraystretch}{1.1}
\begin{tabular}{@{\extracolsep{2pt}}lcccccccccccccccc}
\toprule
Variable & \textit{M} & \textit{SD} & 1 & 2 & 3 & 4 & 5 & 6 & 7 & 8 & 9 & 10 & 11 & 12 & 13 \\
\midrule
1. Idea Quality & $2.85$ & $0.79$ & --- & & & & & & & & & & & & \\
& & & & & & & & & & & & & & & \\

2. Perceived Idea Ownership & $5.42$ & $1.27$ & $.02$ & --- & & & & & & & & & & & \\
& & & & & & & & & & & & & & & \\

3. Idea Diversity & $0.35$ & $0.10$ & $-.26^{**}$ & $.17^{*}$ & --- & & & & & & & & & & \\
& & & & & & & & & & & & & & & \\

4. Creative Self-Efficacy & $4.78$ & $1.41$ & $.10^{*}$ & $.62^{**}$ & $.01$ & --- & & & & & & & & & \\
& & & & & & & & & & & & & & & \\

5. Propensity to Trust Technology & $3.68$ & $0.95$ & $.06$ & $.13^{*}$ & $.02$ & $.23^{**}$ & --- & & & & & & & & \\
& & & & & & & & & & & & & & & \\

6. Perceived Cognitive Workload & $53.20$ & $23.19$ & $.04$ & $.17^{*}$ & $-.05$ & $.03$ & $.18^{**}$ & --- & & & & & & & \\
& & & & & & & & & & & & & & & \\

7. Perceived Technology Agency & $3.06$ & $0.74$ & $-.06$ & $-.21^{**}$ & $-.14$ & $-.18^{*}$ & $-.01$ & $.00$ & --- & & & & & & \\
& & & & & & & & & & & & & & & \\

8. Performance Expectancy & $5.49$ & $1.50$ & $0.11$ & $-.11$ & $-.14$ & $.03$ & $.56^{**}$ & $.11$ & $.00$ & --- & & & & & \\
& & & & & & & & & & & & & & & \\

9. Effort Expectancy & $5.94$ & $1.07$ & $.07$ & $.13$ & $-.02$ & $.18^{**}$ & $.44^{**}$ & $-.01$ & $-.04$ & $.66^{**}$ & --- & & & & \\
& & & & & & & & & & & & & & & \\

10. Hedonic Motivation & $5.19$ & $1.62$ & $.08$ & $.02$ & $-.04$ & $.19^{**}$ & $.60^{**}$ & .07 & $-.06$ & $.73^{**}$ & $.62^{**}$ & --- & & & \\
& & & & & & & & & & & & & & & \\

11. LLM Experience & $3.22$ & $1.22$ & $.09^{*}$ & $.12^{**}$ & $-.02$ & $.19^{**}$ & $.52^{**}$ & 0.00 & $-.03$ & $.28^{**}$ & $.20^{**}$ & $.31^{**}$ & --- & & \\
& & & & & & & & & & & & & & & \\

12. Age & $41.34$ & $12.99$ & $.02$ & $-.01$ & $.04$ & $-.05$ & $.04$ & $.23^{**}$ & $-.05$ &  $.04$ & $-.09$ & $-.01$ & $-.17^{**}$ & --- & \\
& & & & & & & & & & & & & & & \\

13. Total Study Time (Seconds) & $493.33$ & $418.94$ & $.02$ & $-.05$ & $.10^{*}$ & .05 & .08 & $0.18^{**}$ & $.00$ & $.09$ & $.01$ & $.04$ & $.00$ & $.21^{**}$ & --- \\
\bottomrule
\end{tabular}

    \bigskip
    \raggedright\footnotesize
    \emph{Note:} \emph{M} and \emph{SD} represent mean and standard deviation, respectively.
    * indicates \emph{p} $< .05$; ** indicates \emph{p} $< .01$.
  \end{minipage}
\end{table*}

\section{Rater Instructions: Idea Quality}
\emph{Adapted from \citet{lee_empirical_2024}.}

\subsection*{1. Purpose}
You will evaluate ideas that repurpose existing vehicle features (cameras, sensors, screens, lights, or speakers) for entirely new uses.  
Your ratings help us measure the creative quality of each idea.

\subsection*{2. Rating Dimensions}
Use the 7-point Likert scale below for each idea on both dimensions.

\begin{table}[h!]
\centering
\begin{tabular}{cl}
\hline
\textbf{Scale} & \textbf{Anchor meaning} \\
\hline
1 & Not at all / Extremely low \\
2 & Very low \\
3 & Low \\
4 & Moderate \\
5 & High \\
6 & Very high \\
7 & Extremely high \\
\hline
\end{tabular}
\end{table}

\paragraph{Originality / Creativity}  
``Rate how original or creative this idea is in repurposing existing vehicle features.''  
Consider novelty, surprise, and departure from existing uses.

\paragraph{Usefulness / Problem-Solving Potential}  
``Rate how useful this idea would be for solving a meaningful user problem or need.''  
Consider practical value and the extent to which it solves a user need.

\subsection*{3. General Guidelines}
\begin{itemize}
    \item When one person provides several ideas: judge only the first one.
    \item Judge each idea independently. Do not compare it with other ideas you have seen.
    \item Focus on the idea itself, not the writing style, length, or grammar.
    \item Do not penalize unclear wording unless it prevents judging originality or usefulness.
\end{itemize}

\newpage

\section{Coding Scheme: Elaboration Depth}

\label{sec:elaboration-coding}

\begin{table}[htpb]
\caption{Integrated Coding Scheme: Mapping Themes (Levels) to Codes (CIEs)}
\label{tab:coding_scheme}
\centering

\begin{tabularx}{\textwidth}{@{} 
    >{\raggedright\arraybackslash}p{2.8cm}  
    >{\raggedright\arraybackslash}p{3.2cm}  
    X                                       
    >{\raggedright\arraybackslash}p{2.5cm}  
@{}}
\toprule
\textbf{Theme (Level)} & \textbf{Applicable Codes (CIEs)} & \textbf{Description \& Criteria} & \textbf{Depth Move} \\ 
\midrule

\textbf{Level 1}\newline LLM-led Echo & \textit{None} (0 CIEs) & \textbf{Pure Acceptance.} User accepts LLM input as-is. No new characteristics are added. & Endorsement \\ 
\midrule

\textbf{Level 2}\newline Minor Elaboration & \textbf{Feature}\newline \textbf{Mechanism}\newline \textbf{Constraint}\newline \textbf{Audience/Context} & \textbf{Integration of AI details.} User adds codes ($\ge$1), but these details were \textit{already provided by the AI}. Core frame remains unchanged. & Surface Tweak \\ 
\midrule

\textbf{Level 3}\newline Substantive Elaboration & \textbf{Feature}\newline \textbf{Mechanism}\newline \textbf{Constraint}\newline \textbf{Audience/Context}\newline \textbf{Principle} & \textbf{Independent Detailing.} User independently adds codes ($\ge$1) that were \textit{not} present in the AI input. Adds concrete conceptual detailing. & Conceptual Detailing \\ 
\midrule

\textbf{Level 4}\newline Integrative Advancement & \textbf{Integration / Architecture} & \textbf{Incremental New Idea.} User combines, syncs, or restructures components to extend the same core concept based on LLM input. & Integration + Creation \\ 
\midrule

\textbf{Level 5}\newline Human-led Reframing & \textbf{Reframing / Pivot} & \textbf{Radical Innovation.} User replaces the core frame or value logic so the concept serves a different primary purpose or audience. & Conceptual Pivot \\ 
\bottomrule
\end{tabularx} 
\smallskip

\raggedright\footnotesize
\textit{Note.} The coding process followed a two-step logic: First, we analyzed every user interaction to identify the presence of any of the seven codes (CIEs), such as adding a Feature, Mechanism, or Constraint. Second, we assigned a thematic Level (1--5) based on the assigned \textbf{codes} and their \textbf{source} (AI vs. Human).

\end{table}

\color{black}

\section{Perceived Cognitive Workload (Initial Study)}
\label{sec:workload}

\begin{table}[htpb]
\caption{Contrasts of perceived cognitive workload scores across conditions}
\label{tab:nasatlx}
\centering
\begin{tabular}{lcccccc}
\toprule
Contrast & Estimate & SE & 95\% CI & $t$-ratio & $p$-value & Cohen's $d$ \\
\midrule
\modelbadge{} Model-led vs. \questionbadge{} Question-mode & 19.06 & 3.16 & [12.86, 25.30] & 6.04 & <.0001*** & 0.86 \\
\modelbadge{} Model-led vs. \vanillabadge{} Vanilla        & 10.03 & 3.21 & [3.73, 16.30]  & 3.13 & 0.0037**  & 0.45 \\
\suggestionbadge{} Suggestion-mode vs. \questionbadge{} Question-mode & 17.32 & 3.17 & [11.10, 23.50] & 5.47 & <.0001*** & 0.78 \\
\suggestionbadge{} Suggestion-mode vs. \vanillabadge{} Vanilla       & 8.29  & 3.21 & [1.97, 14.60]  & 2.58 & 0.0102*   & 0.37 \\
\bottomrule
\end{tabular}

\begin{flushleft}
\end{flushleft}
 
\smallskip

\raggedright
\footnotesize
Notes: $p$-values are adjusted using the Holm--Bonferroni method;  reported 95\% confidence intervals are unadjusted; effect sizes are Cohen's $d$. * $p<.05$; ** $p<.01$; *** $p<.001$.

\end{table}

\begin{table}[htpb] 
\centering 
\caption{Linear Regression: Perceived Ownership on Perceived Cognitive Workload} 
\label{tab:regression_ownership_tlx} 
\begin{tabular}{lcccccc}
\toprule
Variable & B & $\beta$ & SE & 95\% CI & $t$-value & $p$ \\
\midrule
Constant   & 4.92  & --    & 0.14 & [4.64, 5.20]   & 34.47 & <.001*** \\
Perceived cognitive workload (NASA-TLX) & 0.01 & 0.17 & 0.002 & [0.005, 0.014] & 3.83  & <.001*** \\
\bottomrule
\end{tabular}

\begin{flushleft}
\end{flushleft}

\smallskip

\raggedright
\footnotesize
Notes: $R^2 = .029$, Adjusted $R^2 = .027$, $F(1,484) = 14.64$, $p < .001$. Standardized ($\beta$) and unstandardized ($B$) regression coefficients are reported. 
SE = standard error; CI = confidence interval. 
$^{*}p<.05$, $^{**}p<.01$, $^{***}p<.001$.

\end{table}

\newpage
\section{Within-Subject Analyses (Initial Study)}
\label{sec:Within_Subject_App}

\begin{table}[h]
  \caption{Fixed Effects from Linear Mixed-Effects Model for Idea Diversity}
  \label{tab:lmm-fixed-effects}

  \begin{minipage}{\columnwidth}
    \centering
    \small

    \begin{tabular}{lccccc}
\toprule
\multicolumn{6}{l}{\textbf{Fixed Effects}} \\
\midrule
Effect & Estimate & SE & \textit{t} & \textit{p} & 95\% CI \\
\midrule
Intercept & 0.366 & 0.009 & 38.61 & $<$.001 & [0.347, 0.384] \\
Model-Led (vs. Question-Mode) & 0.132 & 0.013 & 9.94 & $<$.001 & [0.106, 0.157] \\
Suggestion-Mode (vs. Question-Mode) & 0.001 & 0.013 & 0.09 & .930 & [-0.025, 0.028] \\
Time (Refined -- Initial) & 0.067 & 0.009 & 7.24 & $<$.001 & [0.049, 0.085] \\
Model-Led $\times$ Time & -0.157 & 0.013 & -12.06 & $<$.001 & [-0.182, -0.131] \\
Suggestion-Mode $\times$ Time & -0.093 & 0.013 & -7.08 & $<$.001 & [-0.119, -0.067] \\
\midrule
\multicolumn{6}{l}{\textbf{Random Effects}} \\
\midrule
Component & Variance & SD & & & \\
\midrule
Participant (Intercept) & 0.0045 & 0.067 & & & \\
Residual & 0.0041 & 0.064 & & & \\
\bottomrule
\end{tabular}

    \bigskip
    \raggedright\footnotesize
    \emph{Notes:} Results of linear mixed-effects models predicting idea diversity change
    (refined--initial) as a function of LLM interaction mode (Question-Mode baseline)
    and time. Reported are fixed-effect estimates with standard errors, $t$-values,
    $p$-values, and 95\% confidence intervals, as well as random intercept variance
    components for participants.
  \end{minipage}
\end{table}

\begin{table}[h]
  \caption{Contrasts: Within-Condition Changes in Idea Diversity}
  \label{tab:within-condition-contrasts}

  \begin{minipage}{\columnwidth}
    \centering
    \small

    \begin{tabular}{lccccc}
\toprule
Condition & Change & SE & \textit{t} & \textit{p} & 95\% CI \\
\midrule
Question-mode   &  0.067 & 0.009 &  7.20 & $<$.001*** & [0.049, 0.086] \\
Model-Led       & -0.089 & 0.009 & --9.80 & $<$.001*** & [--0.107, --0.071] \\
Suggestion-mode & -0.026 & 0.009 & --2.78 &  .006**   & [--0.045, --0.008] \\
\bottomrule
\end{tabular}

    \bigskip
    \raggedright\footnotesize
    \emph{Notes:} Estimated mean change scores in diversity (Refined--Initial) by condition.
    Reported are estimated changes with standard errors, $t$-tests, $p$-values, and unadjusted
    95\% confidence intervals. Significance levels: * $p<.05$, ** $p<.01$, *** $p<.001$.
  \end{minipage}
\end{table}

\newpage
\section{Robustness Checks}
\label{sec:RobustnessChecks}

\new{We assessed the assumptions for all regression models across both studies, examining posterior predictive fit, homogeneity of variance, influential observations, and normality of residuals (including random effects for the linear mixed models). While assumptions were largely satisfied, some models showed some evidence of potential residual non-normality or heteroscedasticity. Detailed diagnostic plots for all dependent variables can be found in our GitHub.  To ensure robustness, we supplemented the parametric analyses with non-parametric alternatives: Kruskal–Wallis tests with Dunn's post hoc comparisons for the linear regression models, and ordinal mixed models for the linear mixed models. In the Dunn’s post hoc analyses, we primarily report unadjusted p-values. Additionally, we report holm-adjusted p-values that were corrected across all pairwise comparisons simultaneously (even the comparisons we do not report). Because non-parametric procedures do not allow contrast specification, this yields particularly conservative p-values. Additionally, we applied a logit transformation to the cosine similarity values, which are bounded between 0 and 1, to better satisfy distributional assumptions. Importantly, results from the robustness checks were consistent with those from the parametric hypothesis tests and confirm our main conclusions.}

\subsection{Study 1}

\begin{table}[h]
  \caption{Robustness check: Hypothesis tests for idea quality with non-parametric Kruskal--Wallis test}
  \label{tab:quality-hypotheses-kruskal-Study1}

  \begin{minipage}{\columnwidth}
    \centering
    \footnotesize

    \footnotesize
\begin{tabular}{@{}llcccc@{}}
\toprule
Hyp. & Comparison & $z$ & $r$ & $p$ & $p_{\text{adj}}$ \\ 
\midrule
\multicolumn{6}{l}{\textit{Idea quality}}\\
H1a & \questionbadge{} Question-mode $>$ \controlbadge{} Control & $2.23$ & $0.16$ & $.013^{*}$ & $.090$ \\
H1b & \suggestionbadge{} Suggestion-mode $>$ \controlbadge{} Control & $1.07$ & $0.08$ & $.143$ & $.500$ \\
H1c & \modelbadge{} Model-led $>$ \controlbadge{} Control & $3.88$ & $0.28$ & $<.001^{***}$ & $.001^{**}$ \\
H1d & \vanillabadge{} Vanilla $>$ \controlbadge{} Control & $0.21$ & $0.02$ & $.415$ & $.500$ \\
\bottomrule
\end{tabular}

    \bigskip
    \raggedright\footnotesize
    \textit{Notes:} An omnibus Kruskal--Wallis test revealed significant differences in idea quality across interaction modes
    ($\chi^2(4) = 20.50$, $p < .001$). Pairwise comparisons were conducted using Dunn's post hoc test. Reported are $z$-statistics,
    Holm-adjusted one-sided $p$-values (across all 10 pairwise comparisons), and $r$ effect sizes; only hypothesis-relevant
    comparisons are shown. Significance levels: * $p<.05$, ** $p<.01$, *** $p<.001$. The non-significant adjusted $p$-value for
    idea quality is likely a consequence of correcting for all 10 pairwise comparisons. Only hypothesis-relevant comparisons are reported.
  \end{minipage}
\end{table}

\begin{table*}[h]
  \caption{Robustness check: Hypothesis tests for idea diversity with non-parametric Kruskal--Wallis test}
  \label{tab:diversity-hypotheses-kruskal-Study1}

  \begin{minipage}{\textwidth}
    \centering
    \footnotesize

    \footnotesize
\begin{tabular}{@{}llccccccc@{}}
\toprule
 & & \multicolumn{3}{c}{Parametric test (logit transformed)} & \multicolumn{4}{c}{Dunn's test} \\
\cmidrule(lr){3-5} \cmidrule(lr){6-9}
Hyp. & Comparison & $t$ (df) & $p$ & $d$ & $z$ & $p$ & $p_{\text{adj}}$ & $r$ \\ 
\midrule
\multicolumn{9}{l}{\textit{Idea diversity}}\\
H2a & \questionbadge{} Question-mode $>$ \modelbadge{} Model-led & $5.20\;(481)$ & $<.001^{***}$ & $0.74$ & $5.72$ & $<.001^{***}$ & $<.001^{***}$ & $0.41$ \\
H2b & \questionbadge{} Question-mode $>$ \vanillabadge{} Vanilla & $0.27\;(481)$ & $.791$ & $0.04$ & $0.14$ & $.444$ & $.500$ & $0.01$ \\
H3a & \suggestionbadge{} Suggestion-mode $>$ \modelbadge{} Model-led & $1.45\;(481)$ & $.223$ & $0.21$ & $2.40$ & $.008^{**}$ & $.024^{*}$ & $0.17$ \\
H3b & \suggestionbadge{} Suggestion-mode $>$ \vanillabadge{} Vanilla & $-3.42\;(481)$ & $.999$ & $-0.49$ & $-3.11$ & $.999$ & $.993$ & $-0.23$ \\
\bottomrule
\end{tabular}

    \bigskip
    \raggedright\footnotesize
    \textit{Notes:} An omnibus Kruskal--Wallis test revealed significant differences in idea diversity across interaction modes
    ($\chi^2(4) = 46.11$, $p < .001$). Pairwise comparisons were conducted using Dunn's post hoc test. Reported are $z$-statistics,
    unadjusted and Holm-adjusted one-sided $p$-values (across all 10 pairwise comparisons), and $r$ effect sizes; only hypothesis-relevant
    comparisons are shown. Significance levels: * $p<.05$, ** $p<.01$, *** $p<.001$.
  \end{minipage}
\end{table*}

\begin{table*}[h]
  \caption{Robustness check: Hypothesis tests for perceived idea ownership with non-parametric Kruskal--Wallis test}
  \label{tab:ownership-hypotheses-kruskal-Study1}

  \begin{minipage}{\textwidth}
    \centering
    \footnotesize

    \footnotesize
\begin{tabular}{@{}llccccc@{}}
\toprule
Hyp. & Comparison & $z$ & $p$ & $p_{\text{adj}}$ & $r$ \\ 
\midrule
\multicolumn{6}{l}{\textit{Perceived idea ownership}}\\
H4a & \questionbadge{} Question-mode $>$ \modelbadge{} Model-led & $3.88$ & $<.001^{***}$ & $<.001^{***}$ & $0.28$ \\
H4b & \questionbadge{} Question-mode $>$ \vanillabadge{} Vanilla & $3.42$ & $<.001^{***}$ & $.002^{**}$ & $0.25$ \\
H5a & \suggestionbadge{} Suggestion-mode $>$ \modelbadge{} Model-led & $0.93$ & $.177$ & $.500$ & $0.07$ \\
H5b & \suggestionbadge{} Suggestion-mode $>$ \vanillabadge{} Vanilla & $0.52$ & $.301$ & $.500$ & $0.04$ \\
\bottomrule
\end{tabular}

    \bigskip
    \raggedright\footnotesize
    \textit{Notes:} An omnibus Kruskal--Wallis test revealed significant differences in perceived idea ownership across interaction modes
    ($\chi^2(4) = 61.96$, $p < .001$). Pairwise comparisons were conducted using Dunn's post hoc test. Reported are $z$-statistics,
    unadjusted and Holm-adjusted one-sided $p$-values (across all 10 pairwise comparisons), and $r$ effect sizes; only hypothesis-relevant
    comparisons are shown. Significance levels: * $p<.05$, ** $p<.01$, *** $p<.001$.
  \end{minipage}
\end{table*}

\newpage

\subsection{Study 2}

\begin{table}[h]
  \caption{Robustness check: Hypothesis tests for idea quality with ordinal mixed model}
  \label{tab:quality-hypotheses-clmm}

  \begin{minipage}{\columnwidth}
    \centering
    \footnotesize

    \footnotesize
\begin{tabular}{@{}llccccc@{}}
\toprule
Hyp. & Comparison & Estimate & SE & $z$ & $p$ \\ 
\midrule
\multicolumn{6}{l}{\textit{Idea quality}}\\
H1a & \questionbadge{} Question-mode $>$ \controlbadge{} Control & $0.86$ & $0.16$ & $5.55$ & $<.001^{***}$ \\
H1b & \suggestionbadge{} Suggestion-mode $>$ \controlbadge{} Control & $0.02$ & $0.15$ & $0.13$ & $.899$ \\
H1c & \modelbadge{} M-led (1-shot) $>$ \controlbadge{} Control & $0.87$ & $0.14$ & $6.21$ & $<.001^{***}$ \\
H1d & \miterbadge{} M-led (iter.) $>$ \controlbadge{} Control & $0.68$ & $0.15$ & $4.38$ & $<.001^{***}$ \\
H1e & \vanillabadge{} Vanilla $>$ \controlbadge{} Control & $-0.17$ & $0.16$ & $-1.09$ & $.899$ \\
\bottomrule
\end{tabular}

    \bigskip
    \raggedright\footnotesize
    \emph{Notes:} Pairwise comparisons of interaction modes on idea quality using an ordinal mixed model approach
    (cumulative link mixed model with random intercepts for idea and evaluator). Reported are estimates on the log-odds scale,
    standard errors, $z$-statistics, and Holm-adjusted $p$-values (one-sided, right-tailed tests). Estimates represent the change
    in log-odds of being in a higher quality category compared to the control condition. The overall support for the hypotheses is
    robust and equivalent to the parametric linear mixed model test. Significance levels: * $p<.05$, ** $p<.01$, *** $p<.001$.
  \end{minipage}
\end{table}

\begin{table}[h]
  \caption{Robustness check: Hypothesis tests for idea diversity with logit-transformed values}
  \label{tab:diversity-hypotheses-logit}

  \begin{minipage}{\columnwidth}
    \centering
    \footnotesize

    \footnotesize
\begin{tabular}{@{}llccccccc@{}}
\toprule
 & & \multicolumn{3}{c}{Parametric test (logit transformed)} & \multicolumn{4}{c}{Dunn's test} \\
\cmidrule(lr){3-5} \cmidrule(lr){6-9}
Hyp. & Comparison & $t$ (df) & $p$ & $d$ & $z$ & $p$ & $p_{\text{adj}}$ & $r$ \\ 
\midrule
\multicolumn{9}{l}{\textit{Idea diversity}}\\
H2a & \questionbadge{} Question-mode $>$ \modelbadge{} M-led (1-shot) & $3.34\;(634)$ & $.002^{**}$ & $0.44$ & $3.28$ & $.001^{**}$ & $.004^{**}$ & $0.21$ \\
H2b & \questionbadge{} Question-mode $>$ \miterbadge{} M-led (iter.) & $1.11\;(634)$ & $.396$ & $0.16$ & $0.95$ & $.171$ & $.218$ & $0.07$ \\
H2c & \questionbadge{} Question-mode $>$ \vanillabadge{} Vanilla & $-3.22\;(634)$ & $.999$ & $-0.47$ & $-2.51$ & $.994$ & $.969$ & $-0.18$ \\
H3a & \suggestionbadge{} Suggestion-mode $>$ \modelbadge{} M-led (1-shot) & $8.20\;(634)$ & $<.001^{***}$ & $1.06$ & $7.72$ & $<.001^{***}$ & $<.001^{***}$ & $0.49$ \\
H3b & \suggestionbadge{} Suggestion-mode $>$ \miterbadge{} M-led (iter.) & $5.51\;(634)$ & $<.001^{***}$ & $0.78$ & $4.97$ & $<.001^{***}$ & $<.001^{***}$ & $0.35$ \\
H3c & \suggestionbadge{} Suggestion-mode $>$ \vanillabadge{} Vanilla & $1.12\;(634)$ & $.396$ & $0.16$ & $1.46$ & $.073$ & $.218$ & $0.10$ \\
\bottomrule
\end{tabular}

    \bigskip
    \raggedright\footnotesize
    \textit{Notes:} An omnibus Kruskal--Wallis test revealed significant differences in idea diversity across interaction modes
    ($\chi^2(5) = 117.72$, $p < .001$). Pairwise comparisons were conducted using parametric (linear mixed model) and non-parametric
    (Dunn's test) approaches. For the parametric test, reported are $t$-statistics with degrees of freedom, Holm-adjusted one-sided
    $p$-values (across the 6 contrasts), and Cohen's $d$ effect sizes. For the non-parametric test, reported are $z$-statistics,
    Holm-adjusted one-sided $p$-values (across all 15 pairwise comparisons), and $r$ effect sizes; only hypothesis-relevant comparisons
    are shown. Significance levels: * $p<.05$, ** $p<.01$, *** $p<.001$.
  \end{minipage}
\end{table}

\begin{table}[h]
  \caption{Robustness check: Hypothesis tests for perceived idea ownership with non-parametric Kruskal--Wallis test}
  \label{tab:ownership-hypotheses-kruskal}

  \begin{minipage}{\columnwidth}
    \centering
    \footnotesize

    \footnotesize
\begin{tabular}{@{}llccccc@{}}
\toprule
Hyp. & Comparison & $z$ & $p$ & $p_{\text{adj}}$ & $r$ \\ 
\midrule
\multicolumn{6}{l}{\textit{Perceived idea ownership}}\\
H3a & \questionbadge{} Question-mode $>$ \modelbadge{} M-led (1-shot) & $7.82$ & $<.001^{***}$ & $<.001^{***}$ & $0.56$ \\
H3b & \questionbadge{} Question-mode $>$ \miterbadge{} M-led (iter.) & $3.52$ & $<.001^{***}$ & $.001^{**}$ & $0.22$ \\
H3c & \questionbadge{} Question-mode $>$ \vanillabadge{} Vanilla & $5.57$ & $<.001^{***}$ & $<.001^{***}$ & $0.40$ \\
H4a & \suggestionbadge{} Suggestion-mode $>$ \modelbadge{} M-led (1-shot) & $5.93$ & $<.001^{***}$ & $<.001^{***}$ & $0.42$ \\
H4b & \suggestionbadge{} Suggestion-mode $>$ \miterbadge{} M-led (iter.) & $1.43$ & $.076$ & $.076$ & $0.09$ \\
H4c & \suggestionbadge{} Suggestion-mode $>$ \vanillabadge{} Vanilla & $3.70$ & $<.001^{***}$ & $.001^{**}$ & $0.27$ \\
\bottomrule
\end{tabular}

    \bigskip
    \raggedright\footnotesize
    \textit{Notes:} An omnibus Kruskal--Wallis test revealed significant differences in perceived idea ownership across interaction modes
    ($\chi^2(5) = 130.38$, $p < .001$). Pairwise comparisons were conducted using Dunn's post hoc test. Reported are $z$-statistics,
    Holm-adjusted one-sided $p$-values (across all 15 pairwise comparisons), and $r$ effect sizes; only hypothesis-relevant comparisons
    are shown. Significance levels: * $p<.05$, ** $p<.01$, *** $p<.001$.
  \end{minipage}
\end{table}

\newpage

\section{Analyses Validation Study}
\label{sec:purchaseintent}

\begin{figure*}[h]
  \Description[Purchase intent means with confidence intervals by condition]{Point-and-interval plot of purchase intent mean ratings by condition, with vertical error bars showing ninety-five percent confidence intervals. Conditions on the horizontal axis are question-mode, suggestion-mode, model-led, generation-mode, vanilla, and control. Question-mode and model-led have the highest mean purchase intent values, suggestion-mode has the lowest mean, and generation-mode, vanilla, and control lie in between.}
  \centering
  \includegraphics[width=0.9\linewidth]{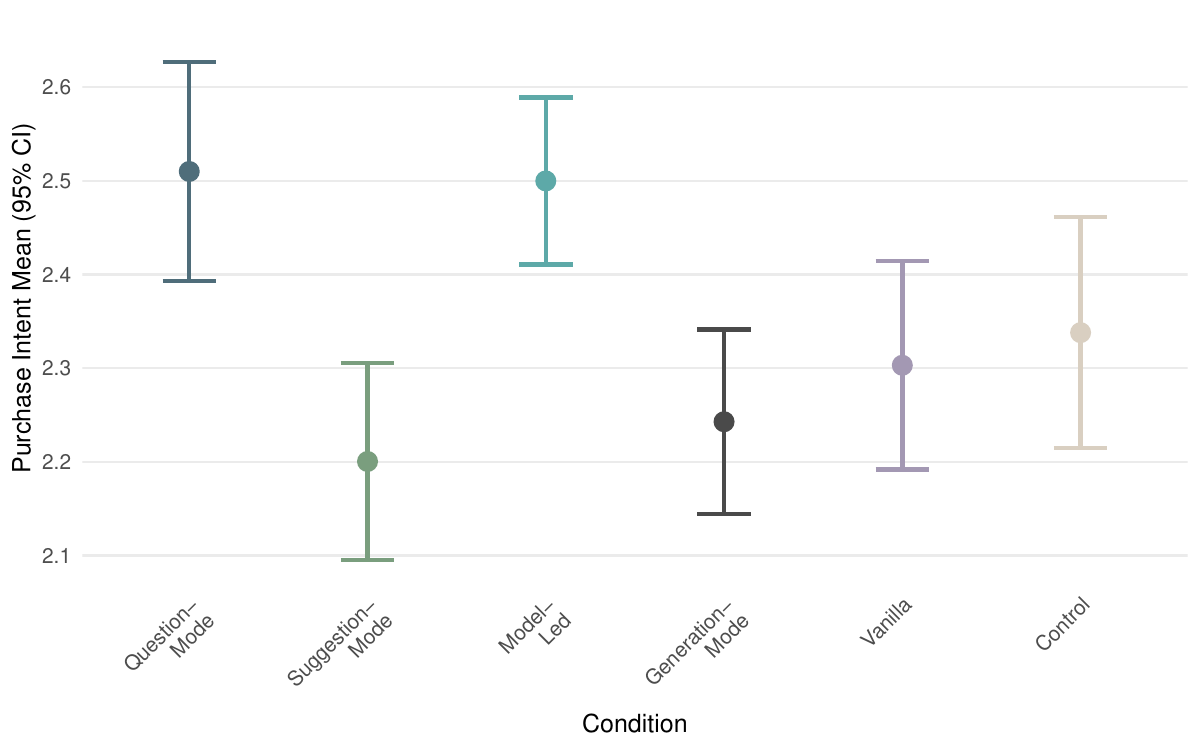}
  \Description{Bar chart showing purchase intent means with 95\% confidence intervals across five conditions. The y-axis ranges from 2.1 to 2.6. Question-Mode shows the highest mean (approximately 2.55), followed by Suggestion-Mode (approximately 2.5), Model-Led Generation-Mode (approximately 2.45), Vanilla (approximately 2.35), and Control (approximately 2.3). Error bars indicate 95\% confidence intervals for each condition, with some overlap between adjacent conditions.}
  \caption{\new{Purchase intent varies across AI interaction modes. Mean purchase intent ratings with 95\% confidence intervals across five conditions (scale: 1--7).}}
  \label{fig:purchase_intent}
\end{figure*}

\begin{figure*}[h]
  \Description[Cognitive workload bars and perceived attribution of creative thinking line]{Combined bar and line chart by condition (question-mode, suggestion-mode, model-led, generation-mode, vanilla, control). Bars show mean cognitive workload on a zero to one hundred scale with error bars. A line with markers shows attribution of creative thinking on a zero to one hundred scale where zero means human did most of the thinking and one hundred means artificial intelligence did most of the thinking. Attribution is highest in generation-mode and lower in question-mode and control. Cognitive workload is higher in the human-led modes and control than in the model-led conditions.}
  \centering
  \includegraphics[width=0.8\textwidth]{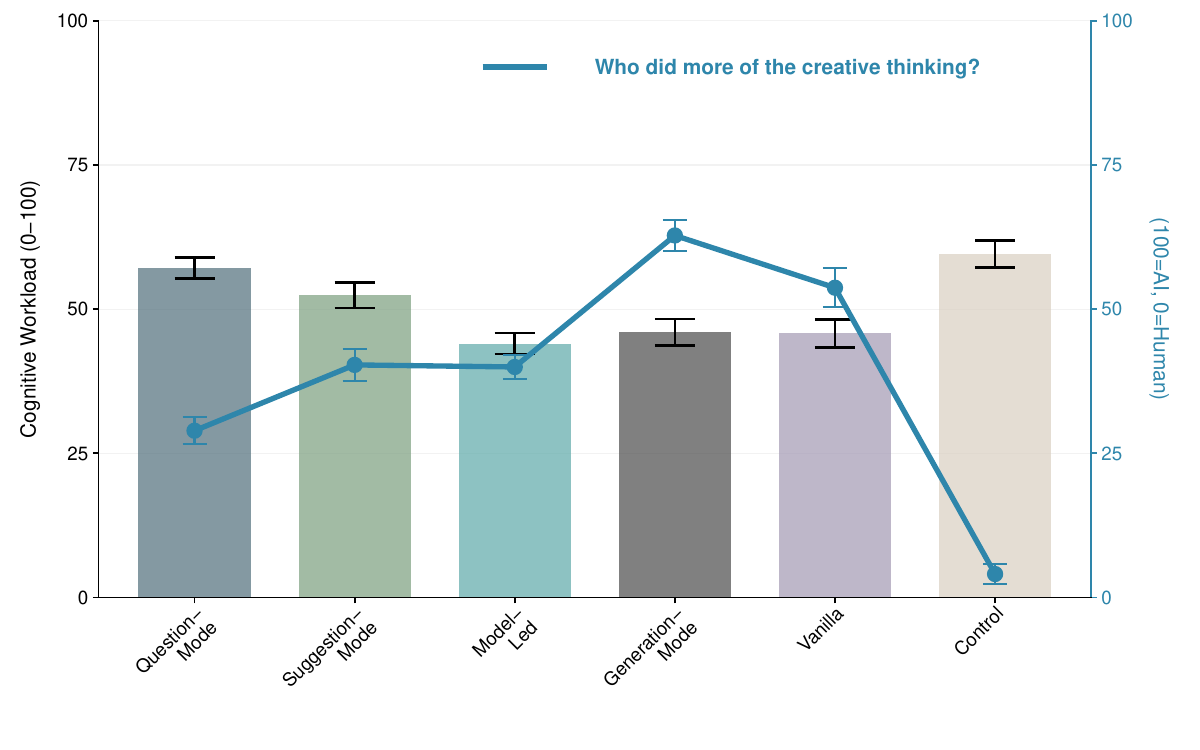}
  \caption{Perceived cognitive workload (bars) and attribution of creative thinking (blue line) across conditions. Cognitive workload was measured using NASA-TLX on a 0--100 scale. Attribution of creative thinking was assessed with the question ``Who did more of the creative thinking required to develop the idea?'' (0 = Human did most of the thinking, 100 = AI did most of the thinking). Error bars represent standard errors. The perceived cognitive workload showed similar patterns to the initial study, with human-led conditions (question-mode $M = 57.14$, $SD = 17.91$; suggestion-mode $M = 52.42$, $SD = 21.87$) showing higher values compared to the model-led conditions (model-led iterative $M = 46.01$, $SD = 23.03$; model-led one-shot $M = 44.04$, $SD = 22.46$), aligning with perceived contribution (human vs.\ AI) except for the model-led one-shot condition ($M = 39.98$, $SD = 25.72$) which showed similar perceived AI contribution to suggestion-mode ($M = 40.34$, $SD = 27.12$).}
  \label{fig:workload}
\end{figure*}

\end{document}